\documentclass[%
 reprint,
 amsmath,amssymb,
 aps,
 prl,
]{revtex4-2}

\usepackage{aas_macros}
\usepackage{graphicx}% Include figure files
\usepackage{dcolumn}% Align table columns on decimal point
\usepackage{bm}% bold math
\usepackage{hyperref}% add hypertext capabilities
\usepackage{color}
\usepackage{natbib}
\usepackage{CJK}
\usepackage[normalem]{ulem}
\usepackage[english]{babel}

\newcommand{\rvw}[1]{#1}

\begin{document}
\begin{CJK*}{UTF8}{gbsn} % Use default fonts from CJK (see below)
\preprint{APS/123-QED}

%\title{A magnetar-powered jet as the origin of blue kilonovae.}

\title{Jets from neutron-star merger remnants and massive blue kilonovae}
%\title{Magnetically-driven outflows from hypermassive neutron star remnants: \\ GRMHD simulations of binary neutron star mergers with weak interactions}
% Force line breaks with \\

\author{Luciano Combi$^{1,2,3}$}
\altaffiliation{CITA National Fellow}

\author{Daniel M.~Siegel$^{4,1,2}$}

\affiliation{$^1$Perimeter Institute for Theoretical Physics, Waterloo, Ontario N2L 2Y5, Canada\\
$^2$Department of Physics, University of Guelph, Guelph, Ontario N1G 2W1, Canada\\
$^3$Instituto Argentino de Radioastronom\'ia (IAR, CCT La Plata, CONICET/CIC)\\
C.C.5, (1984) Villa Elisa, Buenos Aires, Argentina\\
$^4$Institute of Physics, University of Greifswald, D-17489 Greifswald, Germany}

%\collaboration{MUSO Collaboration}


\begin{abstract} 
We perform 
%high-resolution
three-dimensional general-relativistic magnetohydrodynamic simulations with \rvw{weak interactions}
%transport
of binary neutron star (BNS) mergers resulting in a long-lived remnant neutron star, with properties typical of galactic BNS and consistent with those inferred for the first observed BNS merger GW170817. We demonstrate self-consistently that within $\lesssim\!30$\,ms post-merger magnetized ($\sigma\sim 5-10$) \rvw{incipient} jets emerge with asymptotic Lorentz factor $\Gamma\sim 5-10$, which successfully break out from the merger debris within $\lesssim\!20$\,ms. A fast ($v\lesssim 0.6c$), magnetized ($\sigma\sim 0.1$) wind surrounds the jet core and generates a UV/blue kilonova precursor on timescales of hours, similar to the precursor signal due to free neutron decay in fast dynamical ejecta. Post-merger ejecta are quickly dominated by MHD-driven outflows from an accretion disk. We demonstrate that within only 50\,ms post-merger, $\gtrsim 2\times 10^{-2}M_\odot$ of lanthanide-free, quasi-spherical ejecta with velocities $\sim\!0.1-0.2c$ is launched, yielding a kilonova signal consistent with GW170817 on timescales of $\lesssim\!5$\,d.
\end{abstract}

\maketitle
\end{CJK*}

%%%%%%%%% Introduction %%%%%%%%%

\textit{Introduction.---}The astrophysical origin of about half of the elements heavier than iron, created via rapid neutron capture (the r-process), remains an open question 
\cite{cowan2021Origin,siegel2022r}. The first observed binary neutron-star (BNS) merger, detected via gravitational waves \cite{abbott2017GW170817} (GW170817), was followed by quasi-thermal emission \cite{abbott2017Multimessenger,coulter2017Swope,soares-santos2017Electromagnetic} consistent with radioactive heating from the nucleosynthesis of r-process elements in the merger debris---a kilonova \cite{metzger2010electromagnetic}. Based on inferred event rates of BNS mergers, the ejecta yield as well as the photometric and spectroscopic properties \cite{villar2017Combined,kasen2017Origin,smartt2017Kilonova,pian2017Spectroscopic,chornock2017Electromagnetic,tanaka2020Systematic,watson2019identification,kasliwal2022spitzer}, this transient not only provided strong evidence for the production of both light (atomic mass number $A\lesssim 135$) and heavy (lanthanide-bearing; $A\gtrsim 136$) r-process elements in this particular event, but also for BNS mergers being a potentially dominant production site of r-process elements in the Universe.

Whereas the origin of the lanthanide-bearing red emission of the GW170817 kilonova peaking on timescales of a week is naturally explained by magnetohydrodynamically (MHD) driven outflows from a massive, self-regulated, neutrino-cooled accretion disk around a final remnant black hole \cite{siegel2017ThreeDimensional,kasen2017Origin,fernandez2019Longterm,christie2019role,li2021Neutrino}, the origin of the early ($\sim$day) blue and ultraviolet (UV) emission remained more elusive. Both the blue and red GW170817 kilonova emission are difficult to explain by dynamical debris from the collision itself \cite{siegel2019GW170817,metzger2020Kilonovae,radice2020Dynamics} (but see Ref.~\cite{kawaguchi2018Radiative} for a corner case). Several alternative mechanisms for the origin of the blue emission post-merger have been considered, including a combination of magnetically driven and neutrino-driven winds from a remnant NS \cite{metzger2018Magnetar,ciolfi2020Magnetically,mosta2020Magnetar}, additional turbulent viscosity in the remnant NS \cite{fujibayashi2020postmerger, radice2018ViscousDynamical}, spiral waves driven into a post-merger accretion disk by non-axisymmetric modes of a remnant NS \cite{nedora2019Spiralwave}, and outflows from a post-merger accretion disk both around a remnant NS \cite{fahlman2018Hypermassive} and a black hole \cite{miller2019Full}.

Here, we demonstrate by means of self-consistent ab-initio simulations of the merger and post-merger phases that the inferred properties of the blue GW170817 kilonova emission can arise naturally from mass ejection within only $\lesssim\!50$\,ms post-merger due to MHD-driven winds from an accretion disk, aided by non-linear hydrodynamic effects. The emergence of a jet from the remnant NS generates a UV/blue precursor signal and, upon collapse of the remnant, might `seed' an ultra-relativistic jet to generate a short gamma-ray burst.

%%%%%%%%% Computational setup %%%%%%%%%

\textit{Computational setup.---}We solve Einstein's equations coupled to the general-relativistic magnetohydrodynamic (GRMHD) equations with weak interactions using an enhanced version \cite{siegel2018Threedimensional,combi2023grmhd} of the flux-conservative code \texttt{GRHydro} \citep{baiotti2003New, mosta2013GRHydro}, which is part of the \texttt{EinsteinToolkit} \footnote{\url{http://einsteintoolkit.org}} open-source code framework \cite{goodale2003Cactus,schnetter2004Evolutions,thornburg2004Fast,loffler2012Einstein, babiuc2019einstein}. We use the numerical setup of Ref.~\cite{combi2023grmhd} (henceforth CS23) with an atmosphere floor of $\rho_{\rm atm} \sim 5 \times 10^2\,{\rm g}\,{\rm cm}^{-3}$. We implement the recovery of primitive variables of Ref.~\cite{siegel2018Recovery}, which provides support for tabulated nuclear equations of state (EOS), weak interactions, and neutrino radiation via a one-moment approximation of the general-relativistic Boltzmann equation \rvw{supplemented by a leakage scheme}  \cite{radice2016Dynamical,radice2018Binary}.
A fixed Cartesian grid hierarchy composed of six nested refinement boxes is used. The finest mesh covers $\simeq\!76$\,km in diameter with a resolution of $\Delta x=180$\,m. The largest box has an extend of 3000\,km. Reflection symmetry across the orbital plane is employed for computational efficiency. A comparison run without imposing the symmetry at slightly lower resolution shows that our conclusions are not affected by this setup (see CS23). %The GRMHD scheme uses an atmosphere floor of $\rho_{\rm atm} \sim 5 \times 10^2\,{\rm g}\,{\rm cm}^{-3}$. 

The initial data consists of two cold, $\beta$-equilibrated, equal-mass NSs of radius 11.6\,km and mass $1.35M_\odot$ in quasi-circular orbit at a separation of 45\,km. We build initial data with the elliptical solver LORENE \cite{gourgoulhon2001Quasiequilibrium}, employing the APR EOS \cite{akmal1998Equation} in finite-temperature, tabulated form \cite{schneider2019AkmalPandharipandeRavenhall}. This EOS generates cold non-rotating NSs with a maximum mass of $2.2M_\odot$ and a radius of 11.6\,km for a $1.4M_\odot$ NS, within the ballpark of current constraints \cite{landry2020nonparametric,cromartie2020relativistic}. While similar post-merger phenomenology is observed for the other, stiffer EOS configurations of CS23, we focus here on the APR configuration, which results in the longest-lived remnant NS. After setting the hydrodynamical variables, we initialize a weak poloidal magnetic seed field confined to the interior of each star with maximum strength of $B_{\rm max} = 3 \times 10^{15}$\,G at the center and total energy of $10^{48}$\,erg. The initial magnetic field is energetically and dynamically insignificant. 

%%%%%%%%% Magnetic field evolution and jet emergence %%%%%%%%%

\begin{figure}[tb!]
  \centering
  \includegraphics[width=\columnwidth]{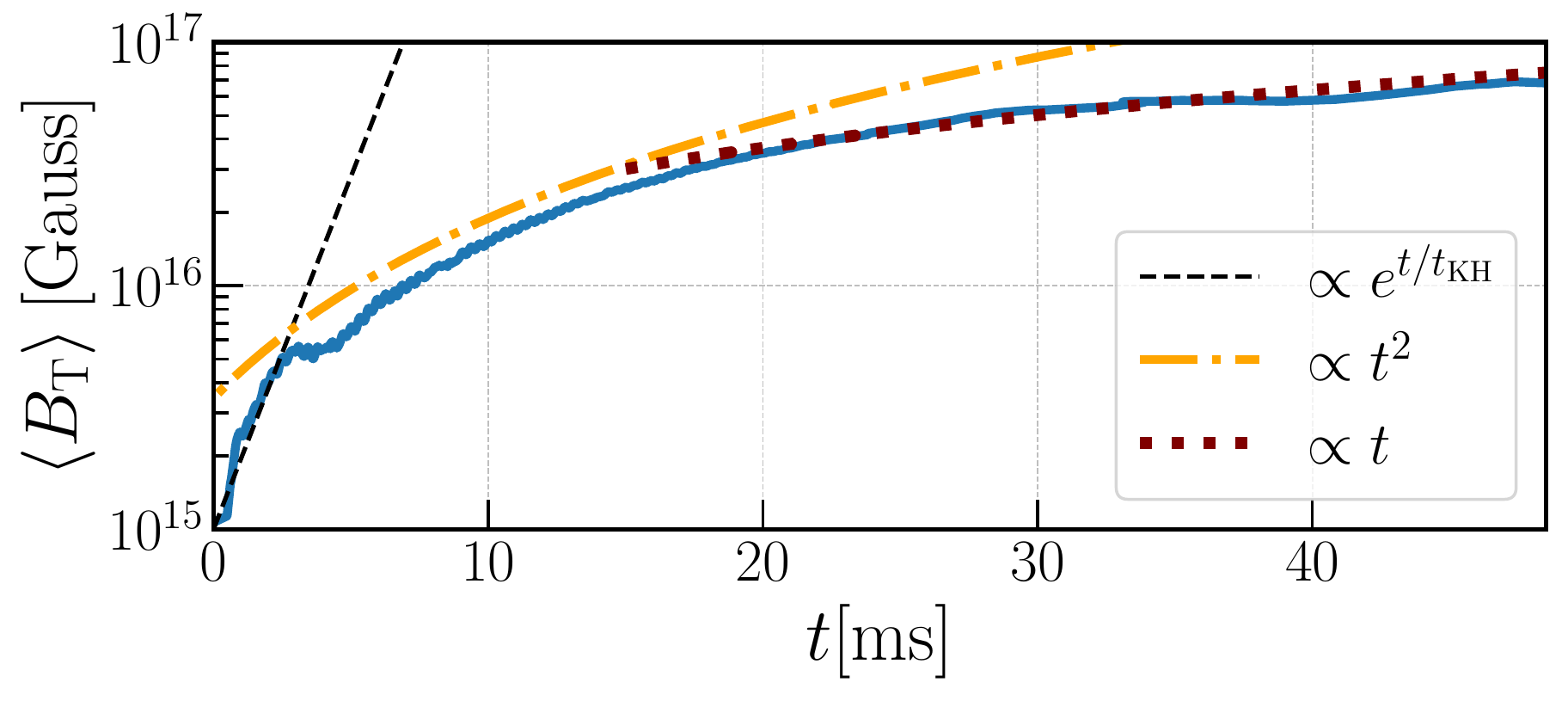}
  \includegraphics[width=0.8\columnwidth]{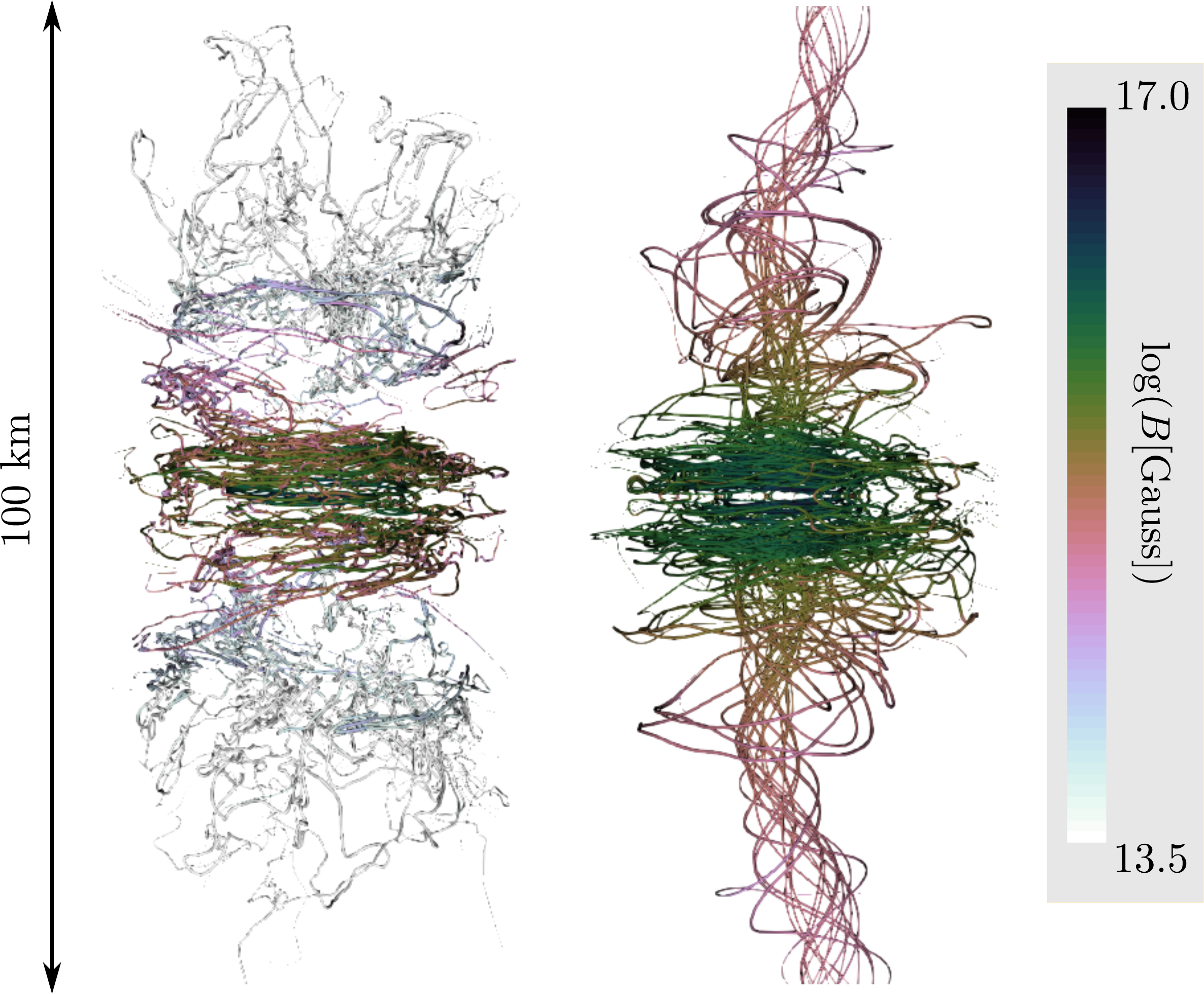}
  \caption{Top: Average toroidal field in the remnant NS as a function of time after merger (blue solid line). The dashed line represents the expected amplification by the Kelvin-Helmholtz instability, the dot-dashed line the turbulent amplification by large-scale vortices, and the dotted line the expected linear growth by magnetic winding. Bottom: 3D rendering of magnetic field lines at 15\,ms (left) and 50\,ms (right), showing the conversion of turbulent fields into large-scale toroidal structures that give rise to twin jets.} %(a magnetic tower; \cite{Lynden-Bell1996}).}
  \label{fig-bTvst-3db}
\end{figure}

\textit{Magnetic field evolution and jet formation.---}During inspiral, the magnetic seed field remains buried inside the stars. Starting at the merger (referred to as $t=0$\,ms), the field is amplified exponentially by the Kelvin-Helmholtz instability (KHI) \cite{price2006Producing,anderson2008magnetized,kiuchi2015Efficient,kiuchi2018Global} for $\approx\!2$\,ms, as evidenced by the toroidal field in Fig.~\ref{fig-bTvst-3db}. Since the KHI has no characteristic spatial scale, the finite resolution of our simulation is only able to capture partial amplification of the average toroidal magnetic field from essentially zero G to $\approx\!5\times 10^{15}$\,G. In the first $15$\,ms post-merger, the generation of large-scale eddies in the remnant NS's interior by quasi-radial core bounces further amplifies the toroidal field by an order of magnitude and roughly $\propto\!t^2$. These currents are associated with the formation of vortices around the inner core \cite{kastaun2021numerical, kastaun2015Propertiesa, ciolfi2017General} which dissipate kinetic into magnetic energy. 

The toroidal field eventually continues to grow at the expected linear growth rate of magnetic winding due to differential rotation inside the \rvw{remnant NS}. Amplification proceeds mainly in the slowly rotating core with positive angular velocity gradient interior to $r \approx 8$\,km. At larger radii, a nearly rotationally supported `Keplerian envelope' is established that gradually transitions into an accretion disk formed by merger debris, a well-established quasi-universal configuration \cite{kastaun2015Properties, ciolfi2017General, kastaun2021numerical, fujibayashi2018Mass}. \rvw{Higher resolution} would lead to earlier saturation of the turbulent magnetic field via the KHI \citep{chabanov2023crustal, palenzuela2021turbulent, aguilera2023role}, making the subsequent amplification processes obsolete, but unlikely altering the qualitative evolution of the system. 

\begin{figure}[tb!]
  \centering
  \includegraphics[width=0.9\columnwidth]{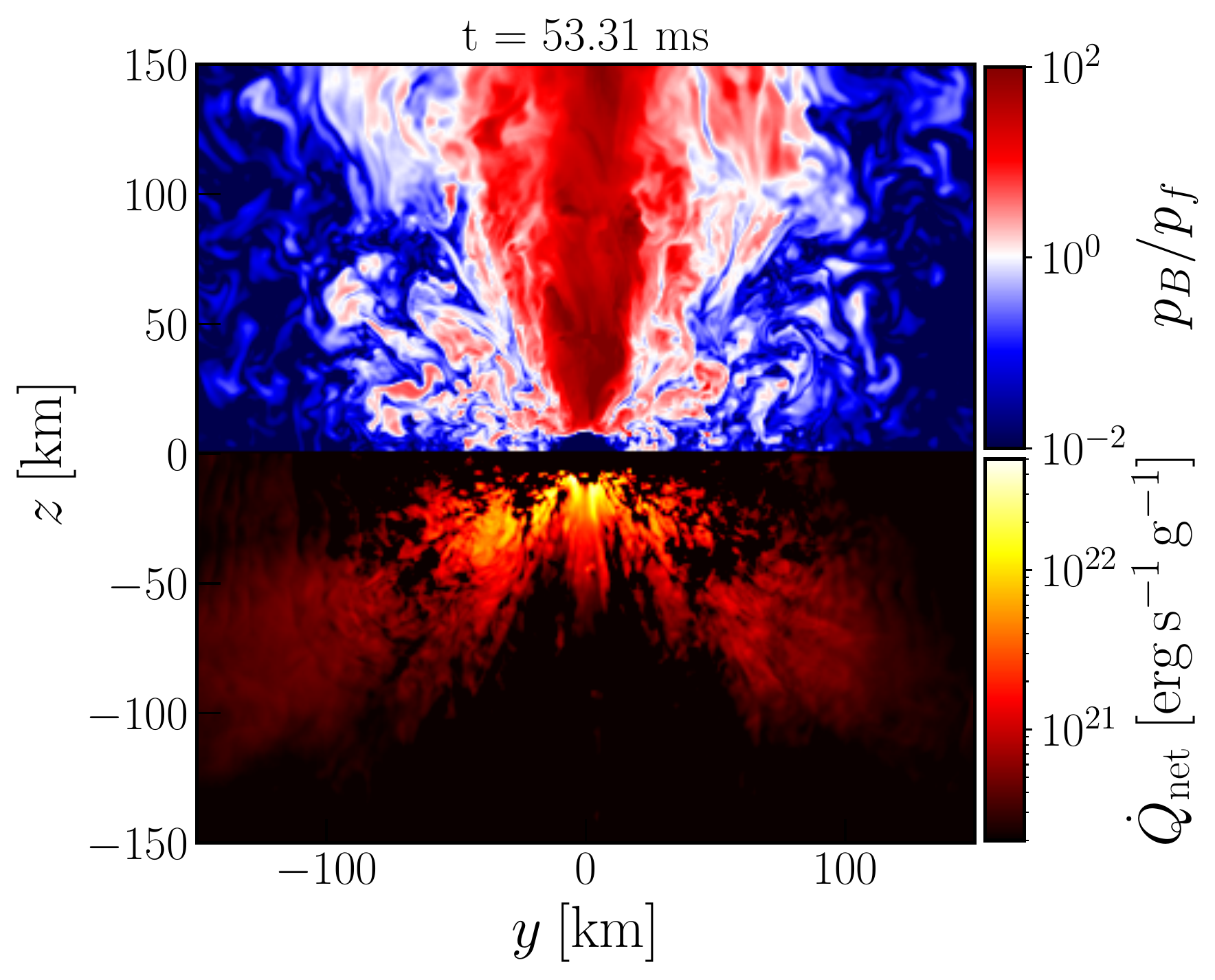}
  \caption{
  %Top: 
  Meridional snapshot showing the magnetic-to-fluid pressure ratio (upper half-plane) and the net specific neutrino heating rate (lower half-plane) once a stationary jet structure has emerged.}
  \label{fig-rho-betaqnet}
\end{figure}

At $\approx\!25$\,ms after merger, thanks to magnetic winding providing an inverse turbulent cascade, the small-scale, turbulent field has been wound up into a large-scale toroidal structure (Fig.~\ref{fig-bTvst-3db}). Owing to their magnetic buoyancy, toroidal fields eventually rise to the stellar surface in the polar regions, break out of the remnant NS, and form a magnetic tower \cite{Lynden-Bell1996} (Figs.~\ref{fig-bTvst-3db}, \ref{fig-rho-betaqnet}). 

\textit{Outflow properties.---}Prior to the emergence of the magnetic tower structure, strong neutrino radiation ($L_{\nu_{\rm e}} \approx 1.7 \times 10^{52}\,{\rm erg}\,{\rm s}^{-1}$) from the hot merger remnant drives a neutrino-driven wind \cite{dessart2009Neutrino,perego2014Neutrinodriven,desai2022three} of unbound material in polar regions with typical velocities $v\lesssim 0.1\,c$ and electron fraction $Y_e\approx 0.5$ via absorption of neutrinos in a gain layer above the stellar surface that extends out to $\lesssim\!50$\,km (Fig.~\ref{fig-rho-betaqnet}). In this gain layer, which is similar to that above proto-NSs in core-collapse supernovae, the net absorbed energy per unit time as seen by the Eulerian observer $\dot{Q}_{\rm net}=\int_{\theta<30^{\circ}} \dot{q}_{\rm net} \rho \Gamma \sqrt{\gamma}\,d^3x$ in a polar volume (polar angle $\theta<30^\circ)$ roughly equals the kinetic power $\dot{E}_{\rm k}= \int_{\theta<30^{\circ}}  \sqrt{\gamma}h \rho \Gamma (\alpha v^i-\beta^i)\,dA_i$ of the wind leaving that volume along the polar axis. Here, $\dot{q}_{\rm net}$ is the net specific neutrino heating rate, $h$ the specific enthalpy, $\Gamma =-n_\mu u^\mu$ is the Lorentz factor of the plasma with three-velocity $v^i$ and four-velocity $u^\mu$, $dA_i$ is the surface element, and $\alpha$, $\mathbf{\beta}$, and $\gamma$ denote the lapse, shift, and determinant of the three-metric of the adopted 3+1 foliation of spacetime with normal vector $n^\nu$ that defines the Eulerian observer. A steady-state wind profile $\rho\propto r^{-2}$ emerges in polar regions, as expected from mass conservation, $\dot{M} = \Delta\Omega\, r^2 \rho \Gamma v$, for a wind opening solid angle $\Delta\Omega$, constant mass-loss rate $\dot{M}$, and four-velocity $\Gamma v$ set by neutrino absorption at the base of the wind. % (Fig.~\ref{fig-rho-betaqnet}).

As the buoyant toroidal magnetic field structures break out of the remnant NS and neutrino absorption helps to form a magnetic tower along the rotational axis, a strongly magnetically dominated outflow with magnetic-to-fluid pressure ratio $p_{\rm B}/p_{\rm f} \approx 10^{2}$ is established (Fig.~\ref{fig-rho-betaqnet}). A steady-state wind profile $\rho\propto r^{-2}$ quickly emerges (Fig.~\ref{fig-rho-betaqnet}), with a steady-state mass-loss rate $\dot{M}$ enhanced by roughly one order of magnitude---consistent with the wind solutions of Ref.~\cite{metzger2018Magnetar}. Within the same time window of 5--10\,ms, the kinetic power of the outflow increases by more than an order of magnitude and comes into equipartition with neutrino heating and the (dominant) Poynting luminosity of the magnetic structure, $\dot{E}_{\rm k}\approx \dot{Q}_{\rm net}+L_{\rm EM}\approx L_{\rm EM}$. The magnetization, $\sigma = L_{\rm EM}/\dot{M}$, in the polar 
%jet
region ($\theta\lesssim 30^\circ$ or specific entropy $s > 25 k_{\rm B}$\,baryon$^{-1}$), rapidly increases to $\sigma \approx 0.1$ (Fig.~\ref{fig-fluxes}). Through strong neutrino heating at the base, shock-heating at the jet head, and in internal shocks, the plasma outflow attains high specific entropy $\gtrsim\!(25-100) k_{\rm B}$\,baryon$^{-1}$. Magneto-centrifugal forces accelerate the plasma in the polar funnel from average velocities of $v \lesssim 0.1c$ (neutrino-driven wind) to $v \approx (0.25-0.6)c$. For acceleration of the wind primarily by magnetic fields, the flow should acquire a four-velocity $ u = v^{r} \Gamma \approx c\sigma^{1/3}$ when it reaches the fast magnetosonic surface \cite{michel1969relativistic}. This limit with the density-averaged value of $\langle u \rangle \approx 0.35c$ is indeed reached in the polar region (Fig.~\ref{fig-fluxes}).

\begin{figure}[tb]
  \centering 
  \includegraphics[width=0.9\columnwidth]{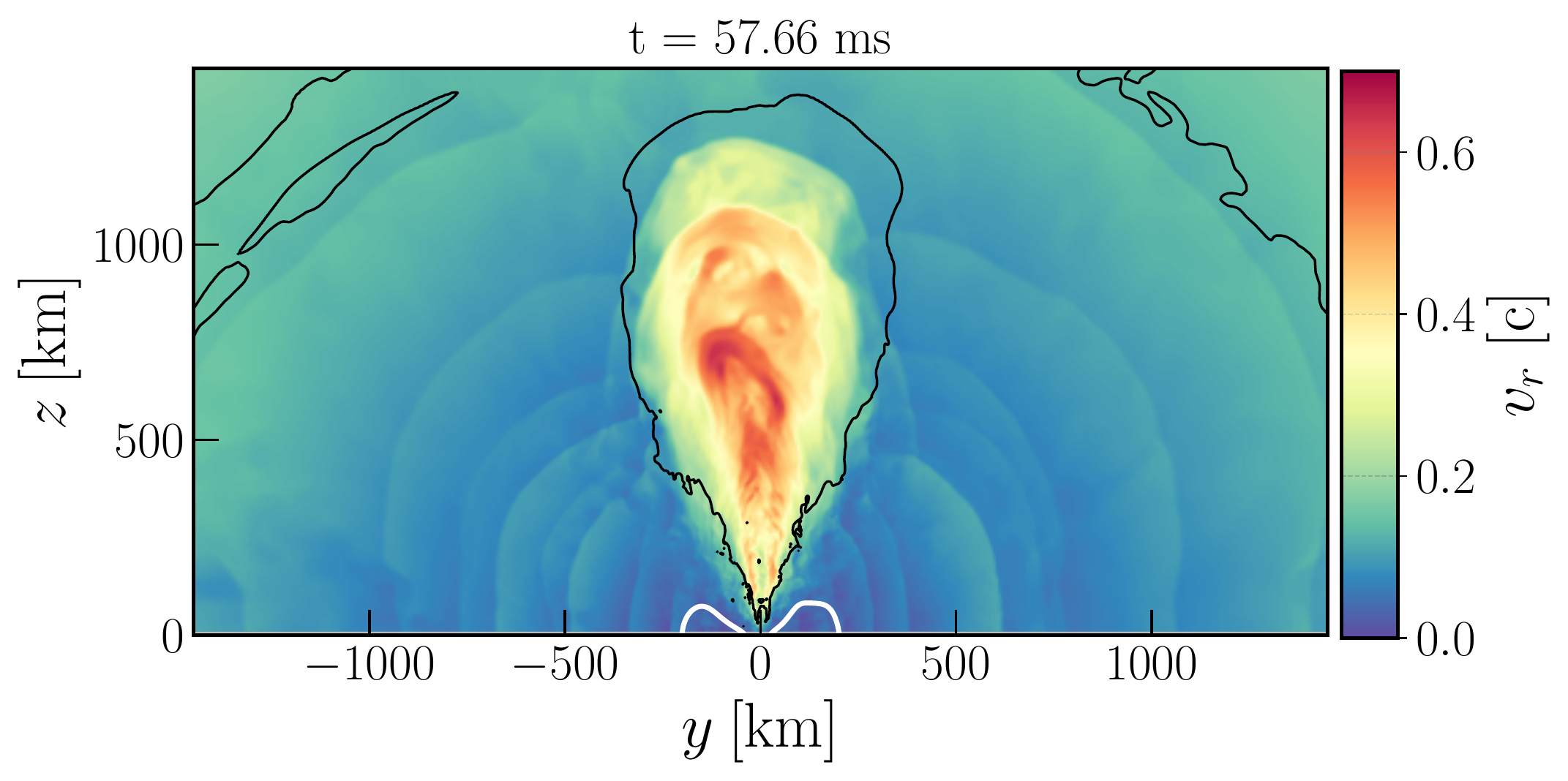}
  \caption{Meridional snapshot of radial velocity showing the successful break-out of a high-velocity, high-entropy jet structure from the surrounding merger debris (dynamical and post-merger wind ejecta; \rvw{SM}). The white contour contains bound material according to the Bernoulli criterion (the accretion disk). Black contours contain dynamically unbound ejecta (geodesic criterion). The rest of the outflow domain is unbound according to the Bernoulli criterion.
  }
  \label{fig-vr-s}
\end{figure}
A magnetized ($\sigma \sim 5-10$) jet structure emerges in the polar funnel consisting of exclusively dynamically unbound material (geodesic criterion; $-u_0 < 1$) with an half-opening angle of $\approx 20^\circ$, and high entropy, $s \gtrsim 50 k_{\rm B}$\,baryon$^{-1}$. The jet head propagates with $v\gtrsim 0.6c$ through and breaks out from the envelope of merger debris (Fig.~\ref{fig-vr-s}, \rvw{Supplemental Material (SM)}). The jet is stabilized by subdominant
%($\sigma\ll 1$)
toroidal magnetic fields in the core, which reduce instabilities at the jet-envelope boundary layer \cite{gottlieb2020structure} while avoiding global (kink) instabilities that can develop in strongly magnetized jets ($\sigma \gg 1$) \cite{bromberg2016relativistic}. The terminal velocity of the jet outflow can be further boosted by neutrino pair-annihilation (expected to be subdominant; not included here) and dissipation of magnetic and thermal energy into kinetic energy at larger spatial scales, up to $\Gamma \lesssim - u_0 (h/h_{\infty} + b^2/\rho) \approx 5-10$. Here, $h_\infty$ denotes the EOS-specific asymptotic value of $h$ and $b$ is the co-moving magnetic field strength.

During the first 50\,ms post-merger, the system ejects $\gtrsim 2\times 10^{-2}M_\odot$ of matter at a time-averaged rate of $\dot{M} \sim 0.5 M_{\odot}\,{\rm s}^{-1}$. \rvw{We quantify unbound ejecta as matter at radii $>300$\,km using the Bernoulli criterion $-(h/h_\infty)u_0 > 1$, where $h_\infty$ includes the approximate average binding energy of the nuclei formed by the r-process \cite{fujibayashi2020postmerger,foucart2021estimating}.} Before the emergence of the jet structure ($t\lesssim 30$\,ms), spiral waves propagating outward in the accretion disk \citep{nedora2019Spiralwave, nedora2021Dynamical} dominate angular momentum transport and mass ejection of the system of $\lesssim\!1 \times 10^{-2} M_{\odot}$ with mass-averaged velocity $\langle v \rangle\approx 0.1 c$. These waves and associated mass ejection, generated mainly by $m=1,2$ non-axisymmetric density modes of the remnant NS through hydrodynamical 
instabilities \cite{lehner2016Instability,radice2016Onearmed, east2016Relativistic}, are visible as concentric waves in the radial velocity (Fig.~\ref{fig-vr-s}), paralleled by oscillations of the unbound mass flux (Fig.~\ref{fig-fluxes}). As the vicinity of the remnant NS becomes strongly magnetized and the jet structure emerges ($t\approx 30$\,ms), the unbound mass flux increases by an order of magnitude to $\dot{M}\approx 1 M_\odot\,{\rm s}^{-1}$ (Fig.~\ref{fig-fluxes}), quickly dominating the total mass unbound post-merger. Within only $\approx 15$\,ms material of $\gtrsim 1\times 10^{-2}M_\odot$ ($\gtrsim 50\%$ of the cumulative post-merger ejecta) is unbound with mass-averaged velocity $\langle v\rangle\approx 0.15 c$. 

As accretion blocks outward radial mass flux in equatorial regions over timescales of interest, the neutrino and magnetically-driven wind from the remnant NS totaling $1 \times 10^{-3}M_\odot$ escapes in polar directions ($ \theta \lesssim \!30^{\circ}$) with only $0.2 \times 10^{-3} M_\odot$ being launched within the jet core. The dominant contribution to ejected material, however, is launched as winds from the accretion disk. Disk winds intensify after $\approx\!30$\,ms when angular momentum transport by spiral waves through the compact bound merger debris, magnetic stresses in the vicinity of the \rvw{remnant NS}, and the onset of MHD turbulence driven by the magnetorotational instability have established and enlarged the accretion disk to a radius of $\gtrsim\!150$\,km with approximate inflow--outflow equilibrium. The onset of strongly enhanced disk winds also coincides with the first cycles of an emerging dynamo as evident from a `butterfly diagram' similar to that obtained in previous work \cite{siegel2018Threedimensional} (SM). Despite intense neutrino irradiation from the remnant, the disk then settles into a self-regulated state of moderate electron degeneracy $\mu_e/k_{\rm B}T\sim 1$ \cite{siegel2017ThreeDimensional,siegel2018Threedimensional}, which implies high neutron-richness of $Y_e\approx 0.1-0.15$ \cite{beloborodov2003nuclear,chen2007Neutrinocooled,siegel2017ThreeDimensional} (SM). The mass averaged $Y_e$ of the disk indeed shifts from $\approx\!0.25$ ($t<30$ ms) to $\approx\!0.15$ as it approaches a quasi-stationary state with an accretion rate of $\gtrsim\! 1 M_\odot\,{\rm s}^{-1}$ and mass of $\approx\!0.19 M_\odot$.

\begin{figure}[tb!]
  \centering
\includegraphics[width=0.8\columnwidth]{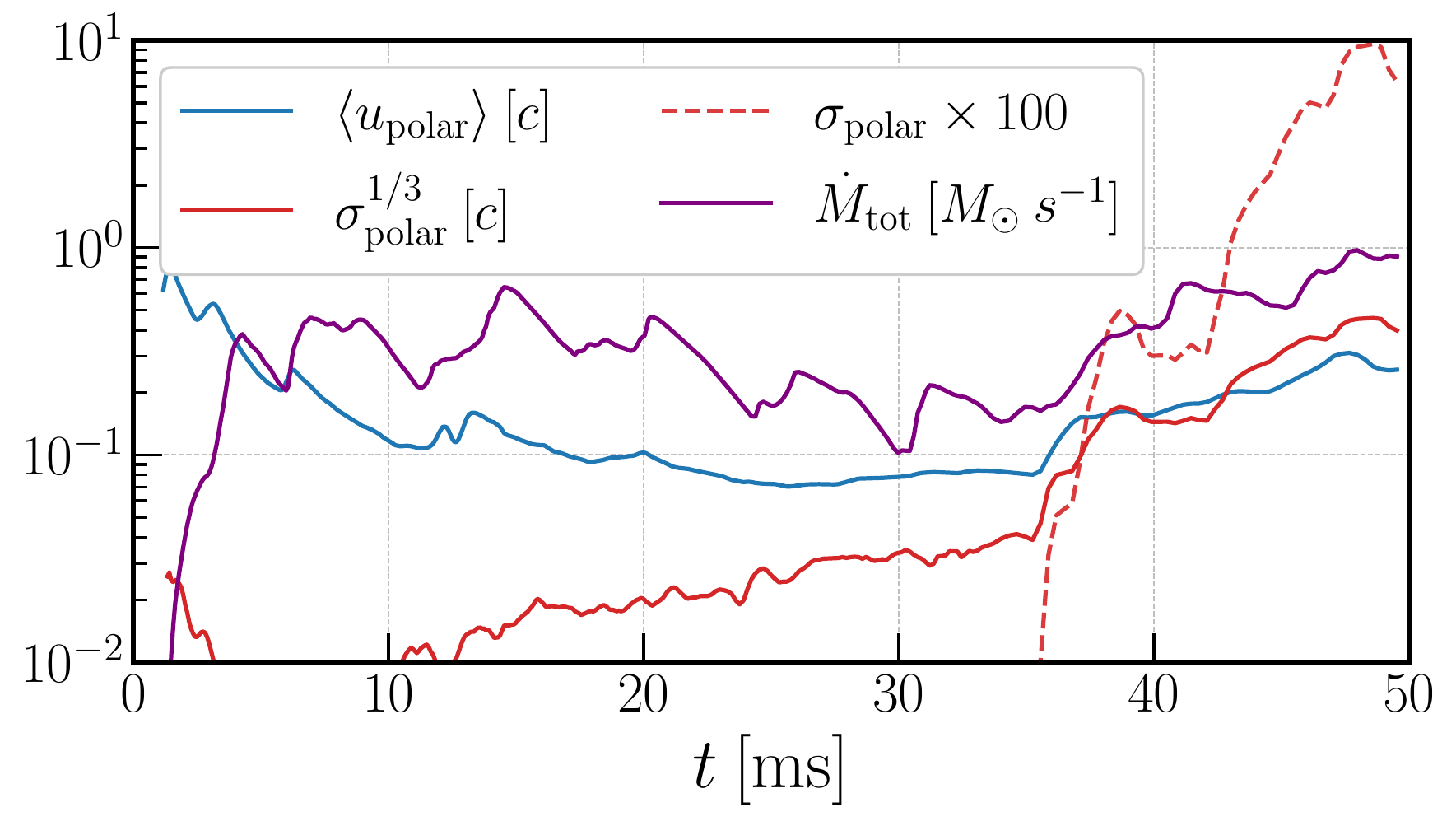}
  \caption{Total unbound mass flux $\dot{M}_{\rm tot}$ through a spherical shell with radius 300\,km and associated density averaged local four-velocity $u =v^r \Gamma$, magnetization $\sigma$, and expected velocity at the magnetosonic surface, $\sigma^{1/3}$, in polar regions ($\theta \lesssim 30^\circ$).}
  \label{fig-fluxes}
\end{figure}
%\begin{figure}[tb]
%  \centering
%\includegraphics[width=0.9\columnwidth]{double_ye_etae_late.pdf}
%  \caption{Meridional snapshot along the rotational axis showing the electron fraction (top) and electron degeneracy $\eta=\mu_e/k_{\rm B} T$ (bottom) with density contours at $\rho = [10^{7.5},10^{8},10^{9.75},10^{13}]\,{\rm g}\,{\rm cm}^{-3}$ as yellow, black, purple, and white solid lines, respectively, $\approx\!50$\,ms post-merger. %}
%  The accretion disk is in a self-regulated state of moderate degeneracy ($\eta\sim 1$), which implies high neutron-richness ($Y_e\approx 0.15$).} % Strong neutrino irradiation from the remnant (cf.~Fig.~\ref{fig-rho-betaqnet}) protonizes the highly neutron-rich disk winds to $Y_e > 0.2$.}
%  \label{fig-ye-eta}
%\end{figure}

%%%%%%%%% Nucleosynthesis \& kilonova %%%%%%%%%

\begin{figure}[tb!]
  \centering
  \includegraphics[width=0.9\columnwidth]{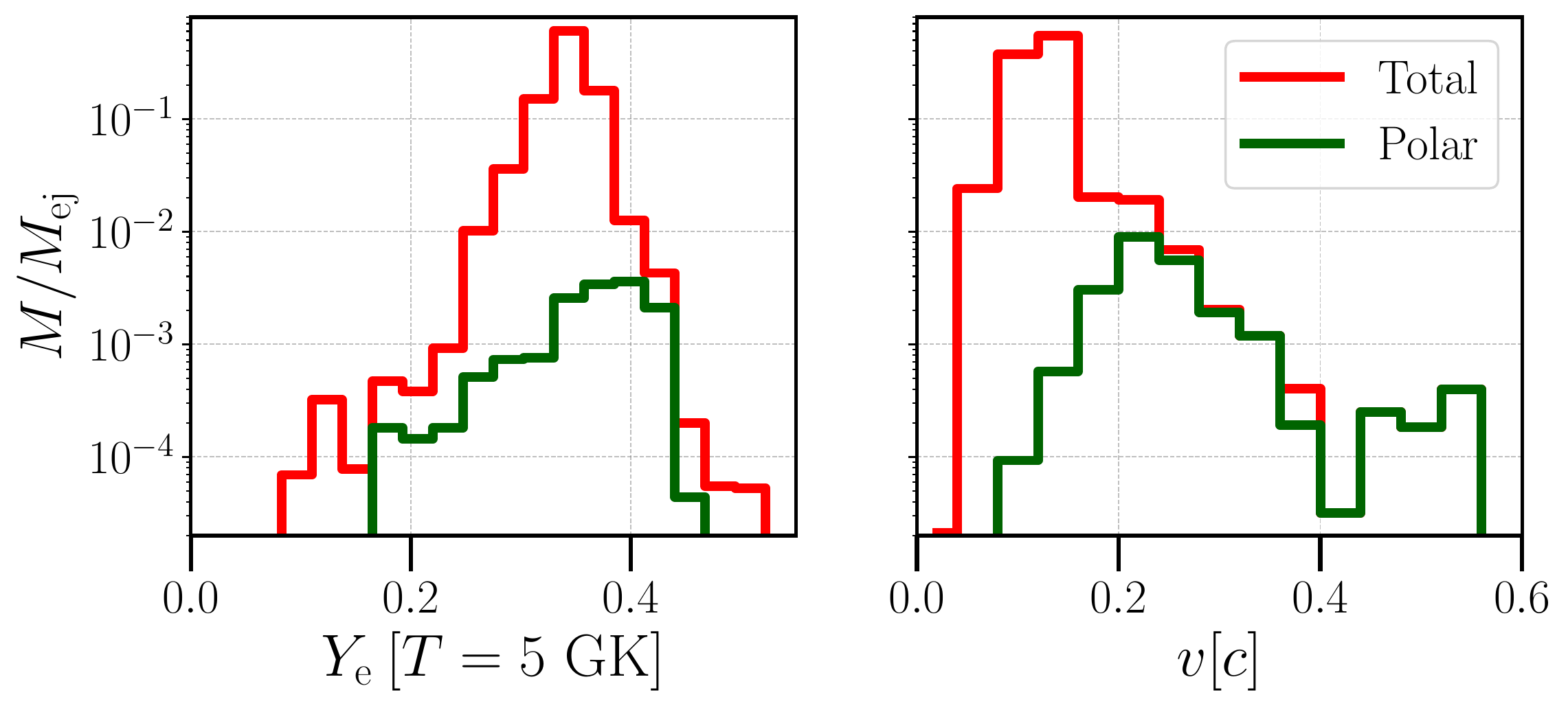}
  \includegraphics[width=0.9\columnwidth]{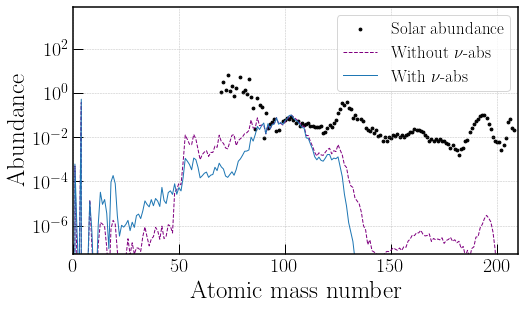}
  \caption{Top: Unbound mass distribution in terms of electron fraction (left; at 5\,GK) and asymptotic expansion velocity (right) as sampled by tracer particles, normalized by the total ejected mass, with separate histrograms for the polar outflows.
  %associated with the jet. 
  %$Y_e$ is extracted at the onset of neutron-capture reactions at 5\,GK. 
  Bottom: Total final nucleosynthetic abundances at $10^9$\,s from reaction network calculations for post-merger ejecta, compared to observed solar abundances \cite{arnould2007Rprocess} (arbitrary normalization) \rvw{with and without including neutrino absorption during r-process nucleosynthesis.}}
  \label{fig-distro}
\end{figure}

\textit{Nucleosynthesis \& kilonova.---}Figure~\ref{fig-distro} shows properties of unbound outflows at the onset of neutron capture reactions ($T\approx\!5$\,GK) as sampled by multiple families of $\approx 2\times 10^{4}$ unbound passive tracer particles injected into the simulation domain (see CS23 for details). Fast outflow speeds $>0.2c$ are almost exclusively associated with polar outflows. 
%As a result of neutrino absorption but high outflow speeds due to magnetic fields, 
Material ejected from the highly neutron-rich degenerate surface layer of the \rvw{remnant NS} is protonized to asymptotic values of $Y_{e} \approx 0.3-0.4$ \rvw{due to neutrino absorption. This is lower than $Y_e\approx 0.5$ as in purely neutrino-driven winds, even in the presence of fast rotation \cite{desai2022three}, due to the accelerating nature of magnetic fields \cite{metzger2018Magnetar}.} %,  much lower than $Y_e\approx 0.5$ as in purely neutrino-driven winds of hot proto-NSs, even in the presence of fast rotation \cite{desai2022three}. 
Outflows from the self-regulated neutron-rich reservoir of the disk are protonized by absorption of intense neutrino radiation from the remnant (cf.~Fig.~\ref{fig-rho-betaqnet}, SM) to a mass averaged value of $\langle Y_e \rangle\approx 0.3$ at 5\,GK.

Nucleosynthesis calculations based on the unbound tracer particles are conducted with the nuclear reaction network \texttt{SkyNet} \cite{lippuner2017SkyNet} using 7843 nuclides and 140 000 nuclear reactions with the setup described in Refs.~\cite{li2021Neutrino} and CS23. They start in nuclear statistical equilibrium at a temperature of $T=7$\,GK and take neutrino irradiation into account using neutrino fluxes directly extracted from our simulation as in Ref.~\cite{li2021Neutrino}. Final abundances at $t=10^9$\,s are shown in Fig.~\ref{fig-distro}. Elements beyond the 2nd r-process peak ($A \approx 130$) are suppressed.

\begin{figure}[tb!]
  \centering
\includegraphics[width=1\columnwidth]{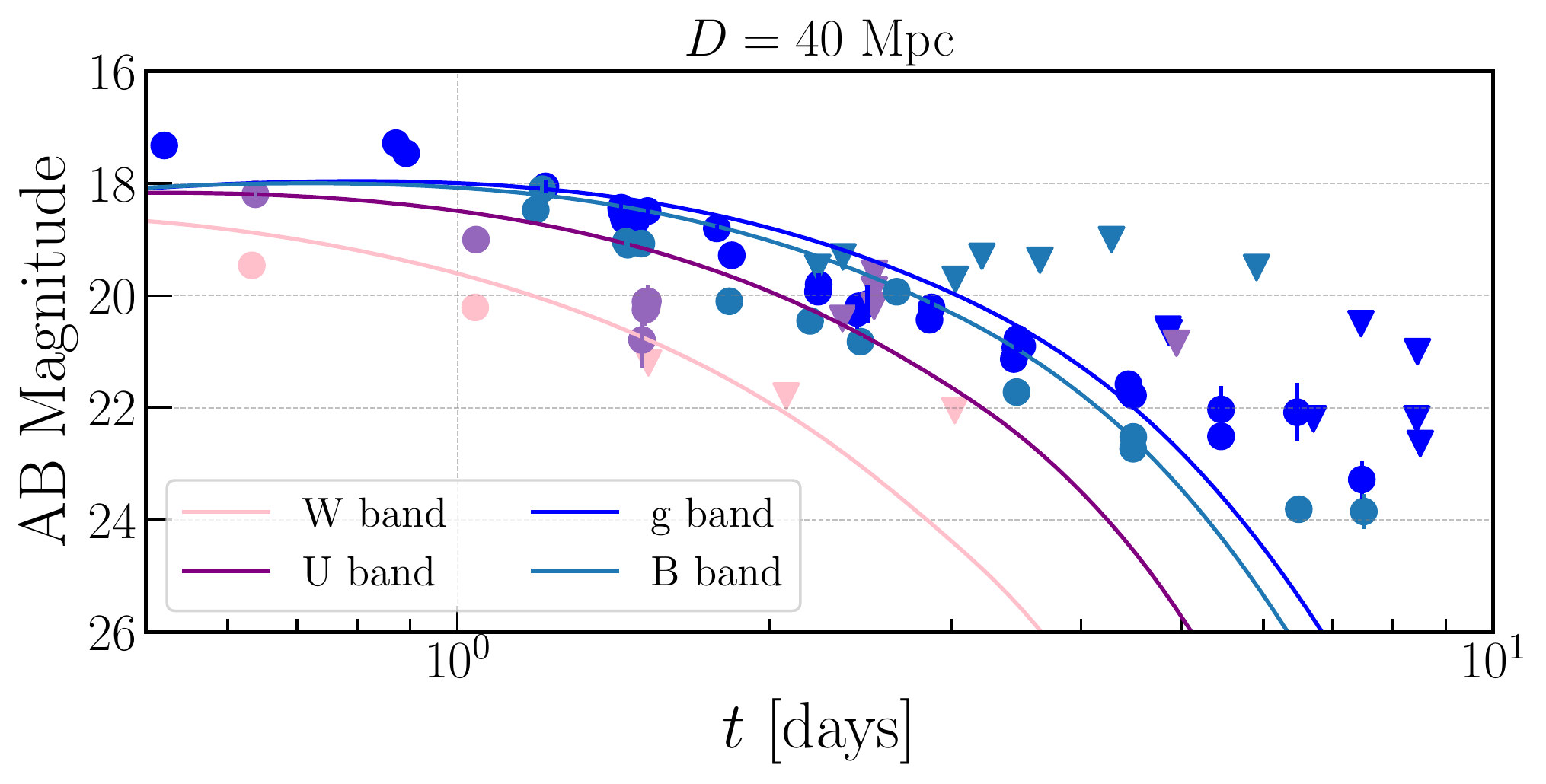}
  \caption{Kilonova light curves of the $\approx\!2\times 10^{-2}M_\odot$ post-merger ejecta in various UV and blue bands, compared to observed data (dots) and upper limits (triangles) of the GW170817 kilonova \cite{villar2017Combined}, and computed for a distance of 40\,Mpc with an observer angle of $35^\circ$ wrt.~the rotational axis, as inferred for GW170817 \cite{abbott2017GW170817, mooley2018superluminal}.
  }
  \label{fig-kn}
\end{figure}

We compute kilonova light curves based on angular-dependent ejecta mass profiles extracted from the simulation, \rvw{using the axisymmetric, viewing-angle dependent model of CS23 (see SM)}. The resulting kilonova signal from post-merger ejecta is consistent with observations of GW170817 in the UV and blue bands up to several days (Fig.~\ref{fig-kn}). Underestimation on timescales $\gtrsim\!5$\,d can be explained by additional `redder' (lanthanide bearing) components \cite{villar2017Combined} not included here, which can be generated by neutron-richer accretion disk winds upon collapse of the remnant into a BH \cite{siegel2017ThreeDimensional,siegel2018Threedimensional,kasen2017Origin,li2021Neutrino}. The $\sim\!{\rm day}$ kilonova is determined by the disk outflows. Fast material from the jet region carries most of the kinetic energy but only 10\% of the total ejected mass; the latter dominates the signal on a few-hours timescale and can boost the $\sim\!{\rm hr}$ UV/blue signal depending on the line of sight (SM). In direction of the jet, the signal is enhanced by up to 1.5 magnitudes, reaching similar luminosities to the kilonova precursor signal from free neutron decay in fast dynamical ejecta \cite{metzger2015Neutronpowered} when the boost due to relativistic effects is taken into account (CS23).

%%%%%%%%% Conclusion %%%%%%%%%

\textit{Conclusion.---}These results provide strong evidence for massive ($\gtrsim 10^{-2}M_\odot$) kilonovae such as GW170817 with early ($\sim\!{\rm day}$) blue and late-time ($\sim\!{\rm week}$) red emission being dominated by post-merger disk outflows. This provides additional support to the conjecture of Ref.~\cite{siegel2022r} that outflows from accretion disks are the main site of the Galactic r-process. We show that binaries consistent with GW170817 and typical of galactic BNS require a remnant lifetime of only $\approx\!50$\,ms to generate a lanthanide-poor blue kilonova component of $\gtrsim\!2\times 10^{-2}\,M_\odot$ with expansion velocity $v\approx 0.15c$ and lightcurves consistent with GW170817. While we find elements of previously proposed mechanisms (magnetized winds \cite{metzger2018Magnetar}, spiral waves \cite{nedora2019Spiralwave}), bulk mass ejection here is due to a combination of \rvw{an incipient} magnetic jet ($\sigma\sim 5-10$), %that emerges $\lesssim 30$\,ms post-merger, 
associated global magnetic stresses, and the onset of MHD turbulence, which reconfigure the accretion disk, enhancing outflows that quickly dominate the cumulative post-merger ejecta. At $\gtrsim\!50$\,ms post-merger the accretion disk with mass of $\approx\!0.19M_\odot$ and accretion rate of $\gtrsim\! 1M_\odot\,{\rm s}^{-1}$ is in a self-regulated neutrino-cooled state with properties in good agreement with initial conditions of previous work \cite{siegel2017ThreeDimensional,siegel2018Threedimensional,li2021Neutrino}. We conclude that upon collapse of the remnant and its neutrino irradiation, lanthanide-bearing outflows of $\gtrsim0.05\,M_\odot$ ($\sim\!30\%$ of the remaining disk mass \cite{siegel2017ThreeDimensional,siegel2018Threedimensional,de2021Igniting,fujibayashi2020postmerger,fernandez2019Longterm,christie2019role}) consistent with the red kilonova of GW170817 are generated over the subsequent few hundred milliseconds \cite{siegel2017ThreeDimensional,siegel2018Threedimensional,li2021Neutrino,fernandez2019Longterm}.

The rapid and self-consistent emergence of a weakly magnetized ($\sigma \sim 0.1$), mildly relativistic ($v\lesssim\!0.6c$) wind from a merger remnant reported here leads to a $\sim\!{\rm hr}$ UV/blue kilonova signal that can be degenerate with the kilonova precursor signal from free neutron decay in the fast tail of dynamical merger ejecta. This novel precursor provides an additional discriminant to distinguish between BNS and NS--black-hole mergers and highlights the importance of early UV/optical follow-up observations of future merger events. The break-out of the jet from the surrounding merger debris observed here may have additional emission signatures, including potential pre-cursors to short gamma-ray bursts \cite{chirenti2023kilohertz}, see SM with Refs. \cite{ciolfi2020Collimated, ciolfi2019First, most2023flares, curtis2023outflows, curtis2023r, radice2021New, kiuchi2023large,hotokezaka2020Radioactive,banerjee2020simulations, wu2021Radiation} for related work. Upon collapse of the remnant NS, the magnetic jet could `seed' the black hole with magnetic flux and forms a strongly magnetized ($\sigma=L_{\rm EM}/\dot{M} \gg 1$), highly-relativistic jet in the absence of stellar winds. This suggests a novel formation mechanism for the central engine of short gamma-ray bursts for remnant lifetimes of $\gtrsim\,30$\,ms.

%%%%%%%%%%%%%%%%%%%%%%%%%%%%%%%%%%%%%%%%%
%% ACKNOWLEDGEMENTS 
%%%%%%%%%%%%%%%%%%%%%%%%%%%%%%%%%%%%%%%%%

\acknowledgments
The authors thank B.~Metzger, D.~Desai, G. Ryan, and E. Most for discussions and comments, and W.~Kastaun for providing visualization tools including the \texttt{PyCactus} package (\href{https://github.com/wokast/PyCactus}{https://github.com/wokast/PyCactus}). This research was enabled in part by support provided by SciNet (www.scinethpc.ca) and Compute Canada (www.computecanada.ca). The authors gratefully acknowledge the computing time granted by the Resource Allocation Board and provided on the supercomputer Lise and Emmy at NHR@ZIB and NHR@G\"ottingen as part of the NHR infrastructure. The calculations for this research were conducted with computing resources under the project mvp00022. LC is a CITA National fellow and acknowledges the support by the Natural Sciences and Engineering Research Council of Canada (NSERC), funding reference DIS-2022-568580. DMS acknowledges the support of the Natural Sciences and Engineering Research Council of Canada (NSERC), funding reference number RGPIN-2019-04684. Research at Perimeter Institute is supported in part by the Government of Canada through the Department of Innovation, Science and Economic Development Canada and by the Province of Ontario through the Ministry of Colleges and Universities.

\section{Supplemental Material}
\subsection{Magnetic field evolution in the remnant and the accretion disk}

\begin{figure}[tb]
  \centering
  \includegraphics[width=\columnwidth]{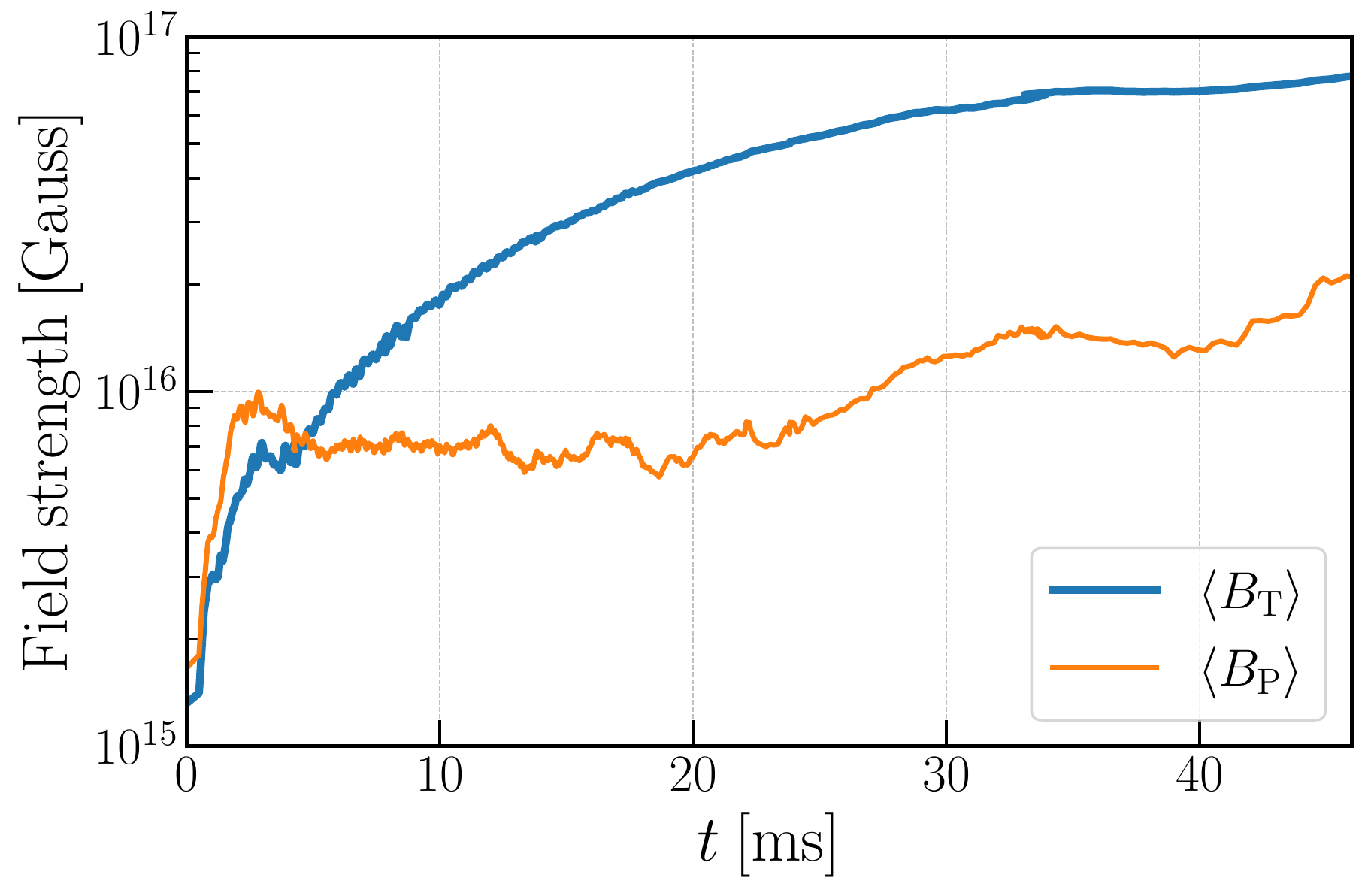}
  \caption{Post-merger evolution of volume averaged poloidal (P) and toroidal (T) magnetic field within the remnant NS (similar to Fig.~1 of the main text).}
  \label{fig-avg-Bs}
\end{figure}

Figure \ref{fig-avg-Bs} illustrates the evolution of the poloidal and toroidal magnetic field components after the merger ($t=0$). During the first $\approx\!2$\,ms post-merger, small-scale turbulence due to the Kelvin-Helmholtz instability (KHI) exponentially amplifies the average toroidal field strength from its initial value of essentially zero G to $5\times 10^{15}$\,G and the average poloidal field strength by at least an order of magnitude to $\approx\!10^{16}$\,G. Whereas the poloidal component has roughly saturated, amplification by large-scale eddies over the subsequent $\approx\!20$\,ms mainly acts on the toroidal component. Magnetic winding starts do dominate magnetic field growth of the average toroidal component starting at $t\approx 25$\,ms. The associated inverse turbulent cascade converts small-scale turbulent fields into large-scale structures that eventually break out of the \rvw{remnant NS} (Fig.~\ref{fig-B-structure} and Fig.~1 of the main text).

\begin{figure}[tb!]
  \centering
\includegraphics[width=1\columnwidth]{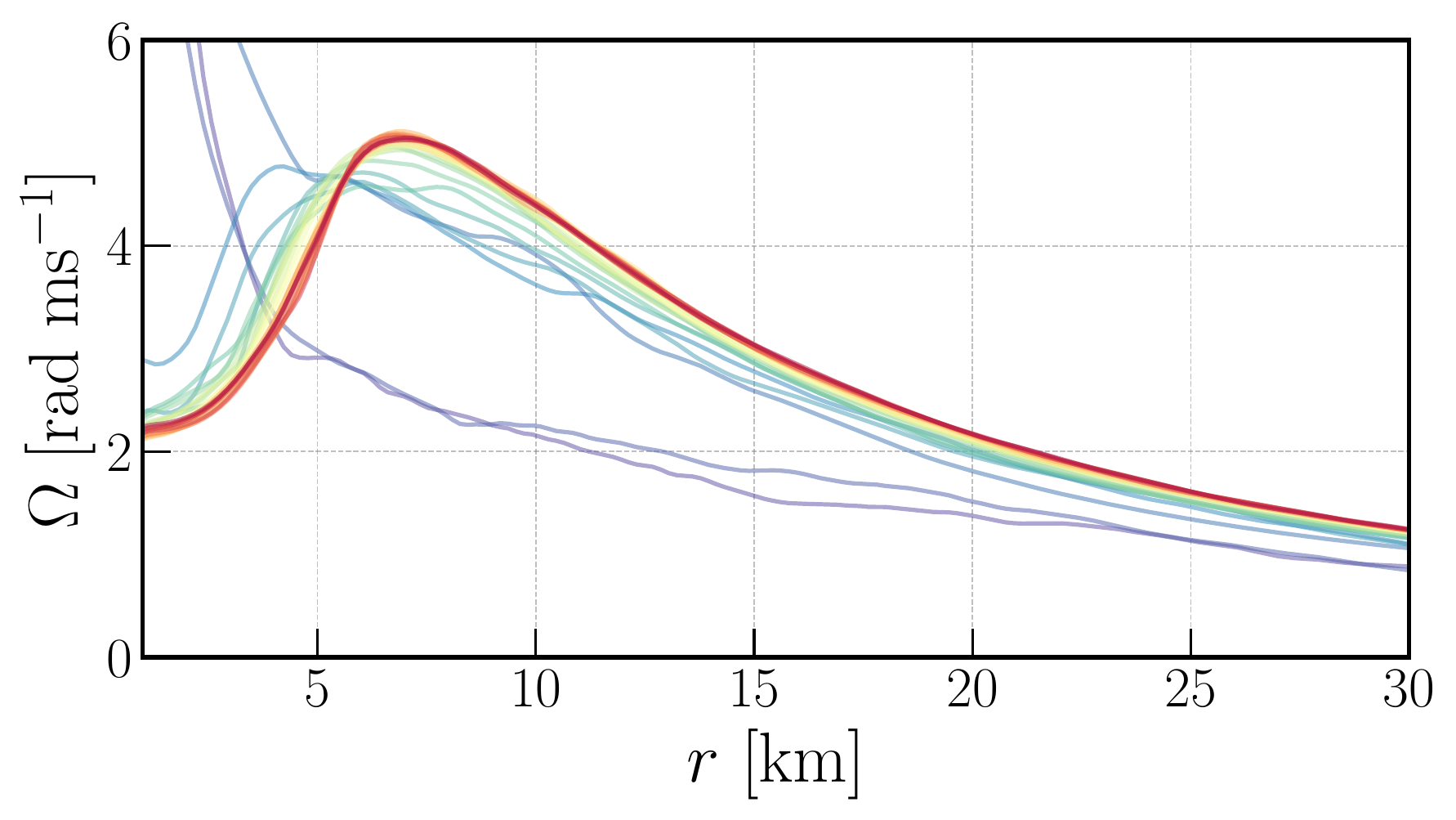}
  \caption{Azimuthally averaged angular velocity of the remnant NS and surrounding accretion disk in the orbital plane at different instances in time (between 11 ms (blue) and 60 \,ms (red)), showing the emergence of a slowly rotating core and a nearly Keplerian envelope with a smooth transition into a rotationally supported accretion disk.}
  \label{fig-omega}
\end{figure}

Initially, post-KHI amplification of the average toroidal field mainly proceeds in the core ($r\lesssim 8$ km, where ${\rm d}\Omega/{\rm d}r > 0$; Fig.~\ref{fig-omega}), and only later spreads into the nearly-Keplerian envelope of the \rvw{remnant NS} ($r\gtrsim 8$ km, where ${\rm d}\Omega/{\rm d}r <0$; see Fig.~\ref{fig-omega}). As the `wound-up' toroidal magnetic field becomes buoyant, it rises toward the stellar surface and eventually breaks out of the \rvw{remnant NS} to form a magnetic tower structure resulting in twin polar jets (cf.~Fig.~\ref{fig-B-structure}). The large-scale structure of the poloidal and toroidal fields after break-out is shown in the lower panel of Fig.~\ref{fig-B-structure}.  The poloidal field in the jet core helps in stabilizing the jet structure against kink instabilities. The jet head propagates at mildly relativistic speeds through the envelope of previously ejected material (dynamical ejecta as discussed in Ref.~\cite{combi2023grmhd} and post-merger disk ejecta) and successfully breaks out from this envelope around $t\approx 50$\,ms. Figure \ref{fig-e-rho} shows the jet once it has successfully drilled through the ejecta structure.

\begin{figure}[tb!]
  \centering
  \includegraphics[width=\columnwidth]{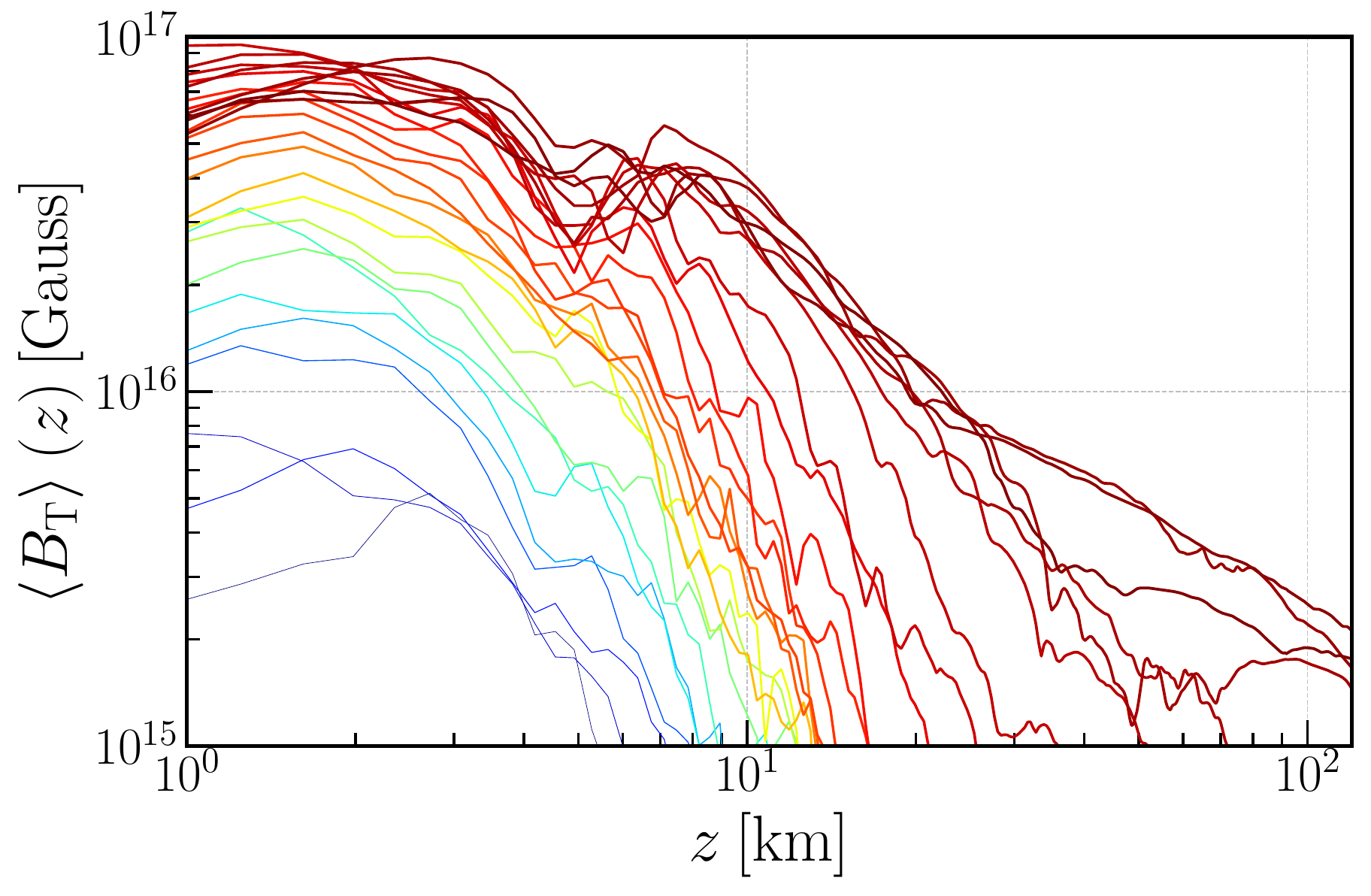}  
  \includegraphics[width=\columnwidth]{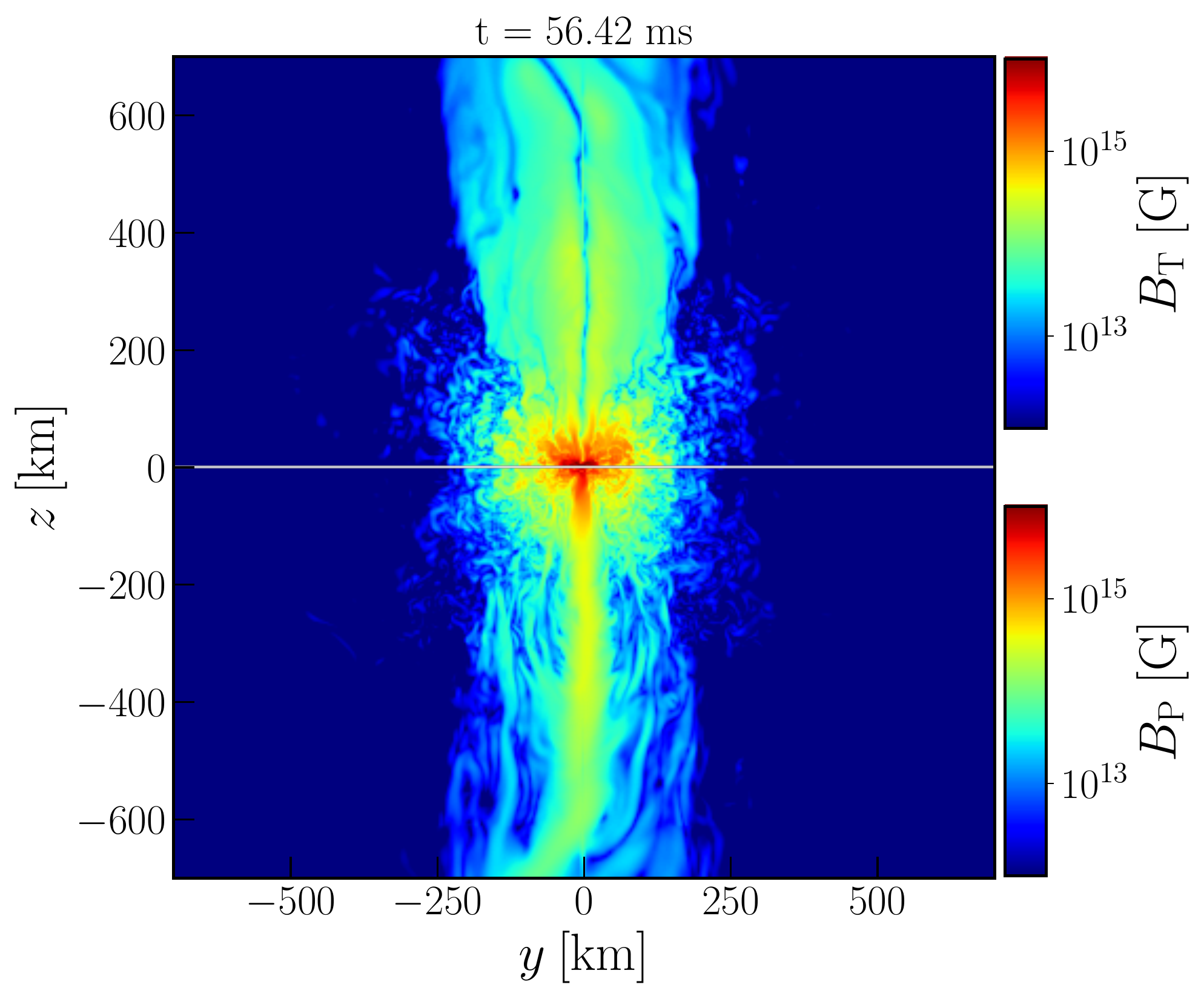}
  \caption{Upper panel: Toroidal magnetic field as a function of height $z$ above the orbital plane, radially averaged within a cylindrical radius $\varpi_{\rm cyl} \le 12$\,km, at early ($t=9$\,ms; blue lines) to late ($t=60$\,ms; red lines) times, with a frequency of $\approx\!1.5$\,ms. The time sequence shows the break-out of the toroidal structures from the stellar surface and the emergence of a $\propto z^{-1}$ magnetic tower structure above the stellar surface ($z\approx 10$\,km). Lower panel: large-scale view of the toroidal (upper half of domain) and poloidal (lower half of domain) magnetic field in the meridional plane at $\approx\!55$\,ms post-merger, showing a mildly relativistic, moderately magnetized ($\sigma \sim 1$) jet structure.}
  \label{fig-B-structure}
\end{figure}

\begin{figure}[tb!]
  \centering
  \includegraphics[width=\columnwidth]{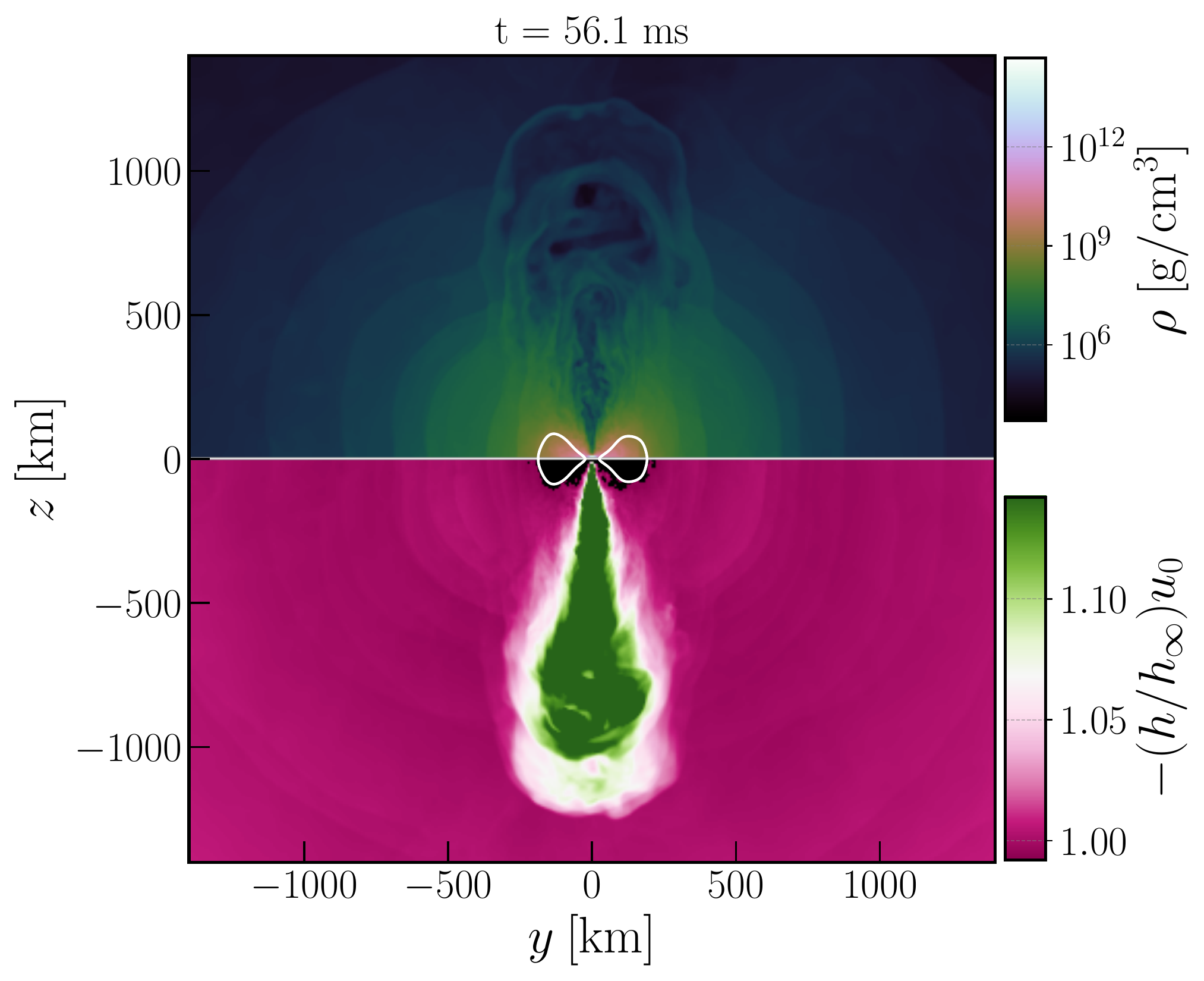}
  \caption{Jet structure in terms of the specific energy (bottom) once it has successfully broken out of the bulk ejecta envelope (represented by rest-mass density; top).}
  \label{fig-e-rho}
\end{figure}

\begin{figure}[ht!]
  \centering
  \includegraphics[width=1.01\columnwidth]{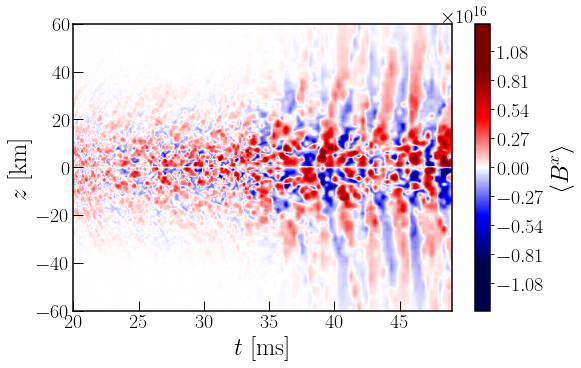}
  \caption{Spacetime diagram of the $x$-component (azimuthal/toroidal component) of the magnetic field in the Eulerian frame, radially averaged between 25 and 60\,km from the rotation axis in the meridional ($yz$) plane, as a function of height $z$ relative to the disk midplane. A stationary dynamo and strongly enhanced mass outflows emerge at around 40\,ms as indicated by the `butterfly' pattern.}
  \label{fig-dynamo}
\end{figure}

\begin{figure}[tb!]
  \centering
  \includegraphics[width=\columnwidth]{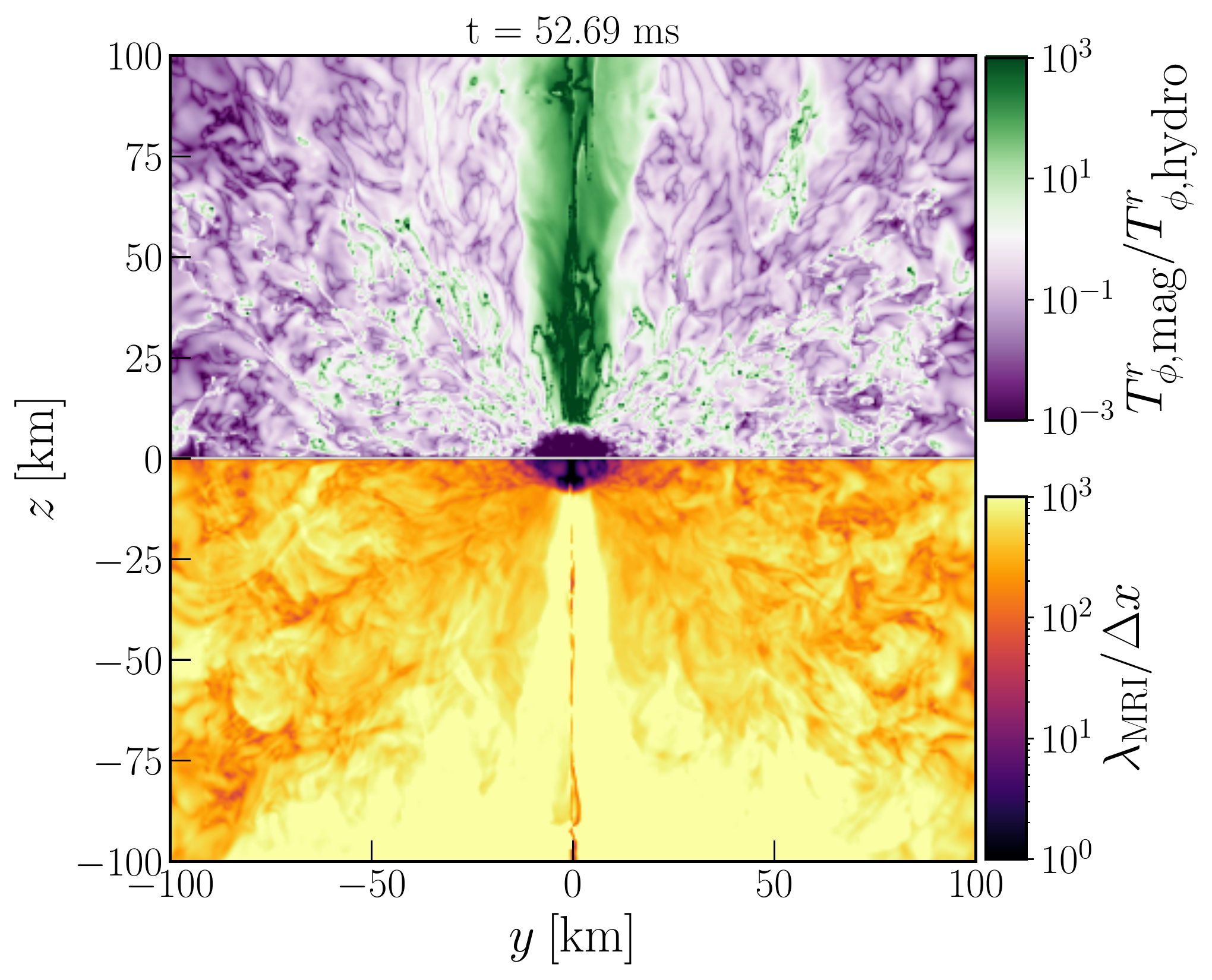}
  \caption{Upper panel: relative strength of total hydrodynamic stress (Reynolds and advective stresses) and magnetic stresses (Maxwell stress) acting on the post-merger plasma at $t\approx 50$\,ms. Lower panel: number of grid points per wavelength $\lambda_{\rm MRI}$ of the fastest-growing unstable MRI mode at the same time instance, indicating that the MRI is well resolved.}
  \label{fig-trphi-lambda}
\end{figure}

\begin{figure}[tb]
  \centering
\includegraphics[width=\columnwidth]{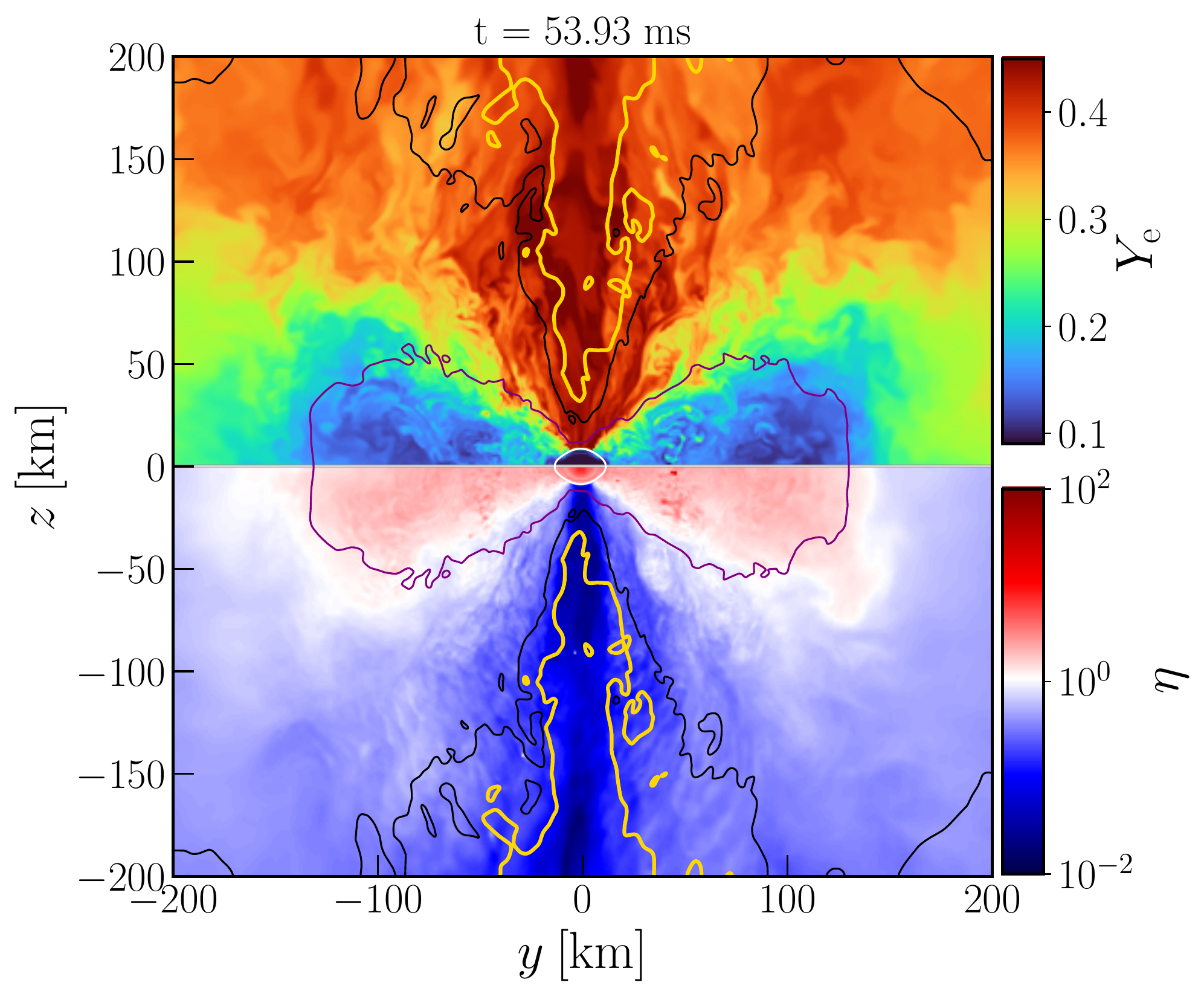}
  \caption{Meridional snapshot along the rotational axis showing the electron fraction (top) and electron degeneracy $\eta=\mu_e/k_{\rm B} T$ (bottom) with density contours at $\rho = [10^{7.5},10^{8},10^{9.75},10^{13}]\,{\rm g}\,{\rm cm}^{-3}$ as yellow, black, purple, and white solid lines, respectively, $\approx\!50$\,ms post-merger. %}
  The accretion disk is in a self-regulated state of moderate degeneracy ($\eta\sim 1$), which implies high neutron-richness ($Y_e\approx 0.15$).} % Strong neutrino irradiation from the remnant (cf.~Fig.~\ref{fig-rho-betaqnet}) protonizes the highly neutron-rich disk winds to $Y_e > 0.2$.}
  \label{fig-ye-eta}
\end{figure}

Magnetic field amplification in the accretion disk that forms from bound merger debris upon circularization around the remnant NS proceeds via turbulent amplification by the magnetorotational instability (MRI). The MRI is well resolved by our numerical setup throughout the MRI-unstable part of the computational domain (${\rm d}\Omega/{\rm d}r < 0$), with typically $\gtrsim 10-100$ grid points per fastest-growing unstable MRI mode (see lower panel of Fig.~\ref{fig-trphi-lambda}). MRI-driven MHD turbulence and associated MHD dynamo activity emerge $\approx\!20$\,ms post-merger once the accretion stream has circularized (Fig.~\ref{fig-dynamo}); it erodes the spiral wave patterns, drives radial spreading of the accretion disk and greatly enhances outflows from the system (see below; Fig.~\ref{fig-outflow}). Magnetic stresses (Maxwell stress) associated with MRI-driven turbulence in the disk and the expulsion of magnetic fields from the remnant NS become comparable to or larger than the total hydrodynamic stresses (Reynolds stress and advective stresses; upper panel of Fig.~\ref{fig-trphi-lambda}) at about 30\,ms, which leads to strong radial spreading and reconfiguration of the accretion disk within only $10-15$\,ms. Fully developed, steady-state MHD turbulence in the disk and an associated dynamo with a cycle of a few ms emerge by $\approx\!40$\,ms as illustrated by the emerging `butterfly' pattern in Fig.~\ref{fig-dynamo}. Figure \ref{fig-ye-eta} shows that the disk then also enters a self-regulated state characterized by moderate electron degeneracy $\eta_e =\mu_e/k_{\rm B}T \sim 1$ and corresponding low $Y_e\approx 0.1-0.15$ \cite{beloborodov2003nuclear,chen2007Neutrinocooled,siegel2017ThreeDimensional,siegel2018Threedimensional}.

\subsection{Comparison with other work}

The post-merger evolution of systems with (meta-)stable remnant NSs as obtained from numerical simulations is sensitive to the microphysics included in the simulations. The inclusion of weak interactions allows the plasma to cool via neutrino emission and thus significantly reduces baryon pollution in the vicinity of the merger remnant immediately after merger. Due to fallback flows becoming less `puffy', neutrino cooling enables the formation of a massive accretion disk around the remnant. Previous work without weak interactions found hot magnetized material surrounding the remnant NS, which inflates rapidly \citep{ciolfi2017General,ciolfi2020Collimated, ciolfi2019First}. The emergence of a large-scale toroidal field then spreads quasi-isotropically, forming a `magnetic bubble' that expands and drives low-velocity winds to large scales \citep{ciolfi2020Magnetically}. Here, neutrino cooling helps to reduce baryon pollution in polar regions and neutrino absorption overcomes the ram pressure of low-angular momentum fallback material in the polar `funnel' and generates a neutrino-driven outflow in polar directions. 
Neutrino absorption and high magnetic pressure help to establish and stabilize a magnetic tower structure. In this wind environment, the jet head is able to propagate and to successfully break out of the merger debris.

\rvw{Ref.~\cite{ciolfi2020Collimated} reports the formation of a collimated outflow powered by a remnant NS, which requires $\gtrsim\!170$\,ms to emerge. Here, we demonstrate by including all relevant microphysical processes that twin incipient jets can emerge by only a few tens of ms post-merger. Furthermore, Ref.~\cite{ciolfi2020Collimated} finds outflows solely in polar directions starting $\gtrsim\!170$\,ms, and they differ quite significantly from our simulations: These outflows are slow with typical velocities of $0.2$c, more similar to our disk outflows, they become unbound only $\gtrsim\!500$\,km above the engine, and they are only loosely collimated as a result of intense baryon pollution.}

Ref.~\cite{mosta2020Magnetar}, place a large-scale dipolar field structure onto a BNS merger remnant and its surroundings (obtained by purely hydrodynamical merger evolution) 17\,ms after merger. The authors find that this imposed global magnetic field structure rapidly leads to the formation of a polar jet-like structure, somewhat similar to that obtained in the present paper. Here, the emergence of twin polar jets is obtained self-consistently by turbulent amplification of an initially vanishing toroidal magnetic field pre-merger and magnetic winding without ad-hoc prescriptions for the magnetic field post-merger. \rvw{After submission of this manuscript, Ref.~\cite{curtis2023outflows} followed the same approach as in Refs.~\cite{curtis2023r, mosta2020Magnetar}, but using a more advanced M1 neutrino transport scheme \cite{radice2021New} instead of a leakage based cooling and heating scheme, finding higher-$Y_e$ outflows and strongly suppressed lanthanides. A central difference of Refs.~\cite{mosta2020Magnetar, curtis2023outflows} to our work is that the former claim polar outflows within the jet to be dominant and, over much longer time spans of hundreds of ms, be consistent with blue kilonovae such as GW170817. The reason for apparent absence of inevitable equatorial outflows due to neutrino-driven winds from the remnant NS and accretion disk winds remains unclear.}

\rvw{At the time of submission of this paper, Ref.~\cite{most2023flares} presented GRMHD simulations with neutrino cooling of BNS mergers using a subgrid model for an $\alpha$-dynamo, which, by construction and depending on the assumed dynamo parameter, can generate an ultra-strong \emph{dominantly poloidal} field of $\gtrsim\!10^{17}$\,G within a few milliseconds after the merger. Strongly amplified magnetic flux structures become buoyant and break out of the remnant NS. Magnetic flux loops anchored to the remnant's surface twist, reconnect, and power flares due to the fact that the remnant rotates differentially. Eventually, a collimated, magnetically dominated outflow forms in the polar directions. %The authors explore whether the associated Poynting luminosity shows similar periodicities to the initial oscillations of the remnant NS. Ref.~\cite{chiarenti} speculates that such oscillations could be responsible for periodicities that they claim are present in the prompt emission of some short GRBs. 
In our simulations, the generation of a jet is based on self-consistent turbulent amplification and winding of a \emph{dominantly toroidal} magnetic field.
%toroidal field is instead amplified by winding, and so it takes a few tens of milliseconds to reach high mangnetizations to break through and form the outflow. Although the velocities are comparable to our simulation, we do not find any periodicities as expected, since the remnant oscillations already died out. Moreover, we do not see flares because of baryon background neutrino driven wind.
}

\rvw{After submission of this paper, Ref.~\cite{aguilera2023role} find with very-high resolution ($\Delta x = 60$\,m) GRMHD simulations using a Large-Eddy subgrid model a dominantly toroidal, turbulently amplified, small-scale magnetic field post-merger, which becomes more coherent due to magnetic winding around $\sim\!30$\,ms post-merger, similar to what we find here. One may speculate that the fact that a large-scale helical, loosely collimated structure emerges from the remnant NS only at late times ($\gtrsim\!100$\,ms post-merger) may be related to the absence of weak interactions and realistic EOSs (cf.~the discussion above related to Ref.~\cite{ciolfi2020Collimated}).}

\rvw{Finally, after submission of this paper, Ref.~\cite{kiuchi2023large} find with ``extremely-high'' resolution ($\Delta x = 12$\,m) GRMHD simulations including weak interactions that a similar toroidally dominated magnetic field and collimated outflow is produced after merger over similar timescales of tens of milliseconds, broadly validating our results. They attribute the formation of a large-scale field to an MRI-driven dynamo acting within the envelope of the star and the inner accretion disk (similar to our Fig.~\ref{fig-dynamo}), and not to the stellar interior as we find here. However, in this proposed MRI scenario, the actual role of the $\alpha$-$\Omega$ dynamo in generating the large-scale flux observed in the simulation remains somewhat uncertain; it remains also unclear what drives a similar growth of toroidal \textit{and} poloidal fields in their low-resolution run, where this MRI-driven dynamo is not resolved.} %, and whether this is independent of the initial magnetic field strengths.}

%\cite{mosta2020Magnetar,curtis2021Process}

%field manage to pull through the evacuated funnel creating a collimated structure. The jet velocity is considerably larger than magnetic winds found in simulations without weak interactions (e.g., \cite{ciolfi2020Magnetically}), but more similar to those of Ref.~\cite{mosta2020Magnetar}, which includes neutrino transport, but assumes a large-scale poloidal magnetic field in the post-merger.

\subsection{Global outflow properties}

\begin{figure}[tb!]
  \centering
  \includegraphics[width=\columnwidth]{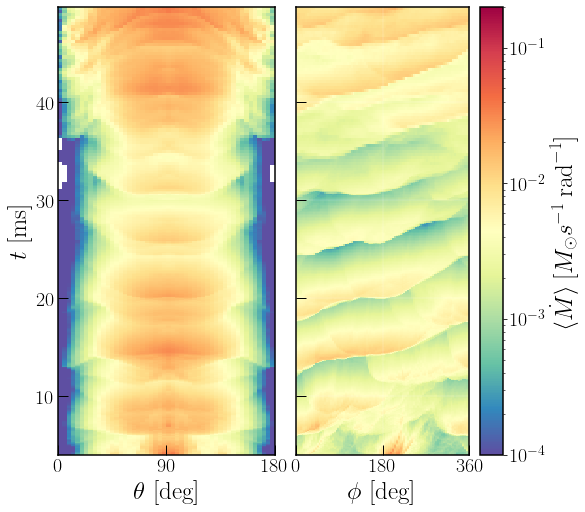}
  \caption{Spacetime diagrams of the mass outflow through a spherical shell with radius 300\,km averaged in the azimuthal (left) and polar (right) direction as a function of time and polar and azimuthal angle, respectively.}
  \label{fig-outflow}
\end{figure}

Spiral waves driven by non-axisymmetric modes in the remnant NS, similar to those observed in previous hydrodynamic simulations \cite{nedora2019Spiralwave,nedora2021Dynamical}, are imprinted in and are associated with the dominant mass outflow during the first $\approx\!30$\,ms post-merger (Fig.~\ref{fig-outflow} and Fig.~5 of the main text). %can be clearly observed in the spacetime plot of Fig.\ref{fig-outflow} measured at radius of 300 km. 
Once the accretion disk spreads radially due to MHD effects and a steady-state dynamo emerges (see above), the outflow enhances drastically (around $t\approx 35$\,ms), quickly dominating the total cumulative mass ejection. Furthermore, the plasma outflow becomes more spherical in this latter phase (cf.~Fig.~\ref{fig-outflow}). %enhancement of mass outflows when the disk settles can also be seen at later times.

\begin{figure}[tb!]
  \centering
\includegraphics[width=1\columnwidth]{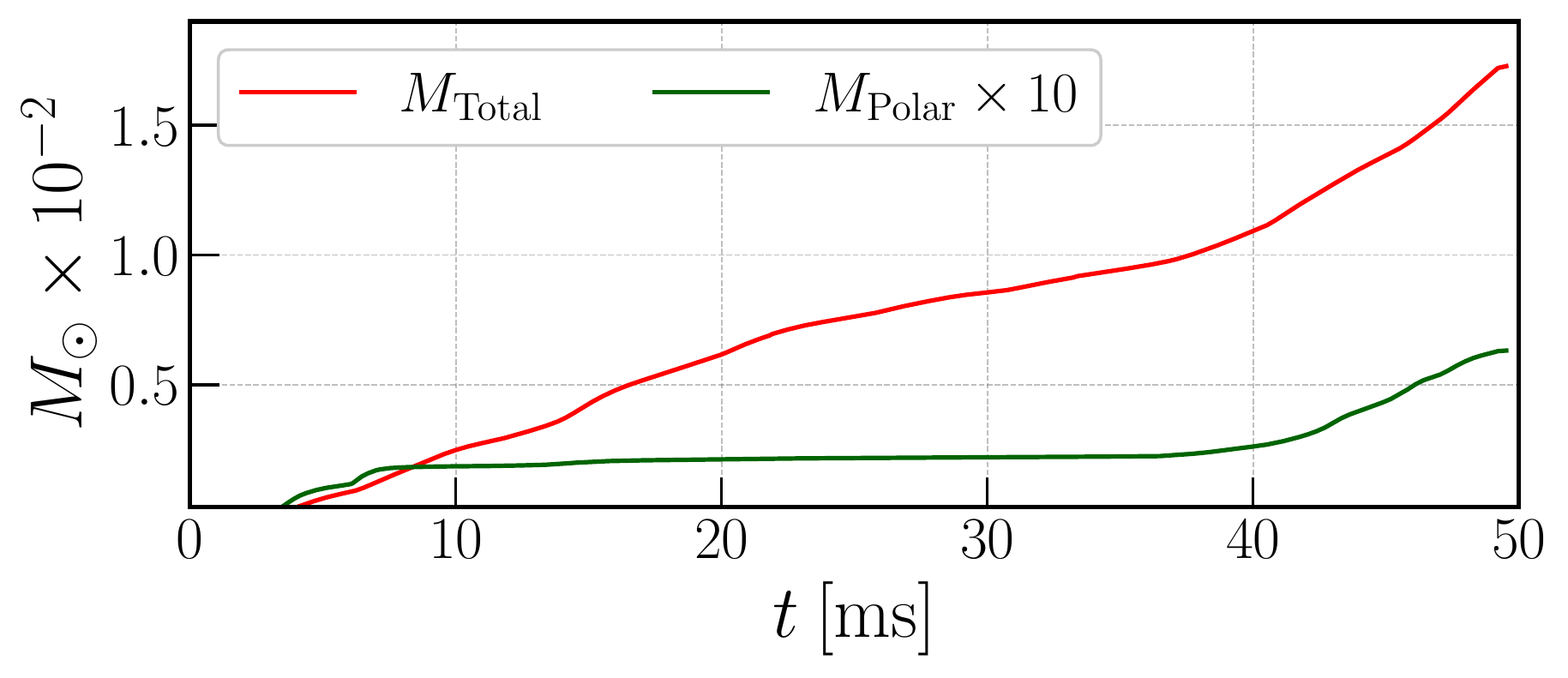}
  \caption{Total (dynamical + post-merger) cumulative ejected mass within the polar regions including the jet ($\theta \lesssim 30^\circ$; green solid line) and total solid angle (jet and disk outflows; red solid line) as a function of time.}
  \label{fig-masses}
\end{figure}

Roughly 95\% of the total cumulative ejected mass of $\gtrsim\!2\times 10^{-2}M_\odot$ by $t\gtrsim 55$\,ms is carried by disk winds (Fig.~\ref{fig-masses}), which have a narrow velocity profile of $v\approx (0.05-0.2)c$ (cf.~Fig.~6 of the main text). The kinetic energy of the outflow, however, is dominated by the polar component including the jet (polar angle $\theta \lesssim 30^\circ$), which extends to asymptotic velocities of $v \gtrsim 0.6c$ (see Fig.~6 of the main text). Only about 1\% of the total outflows originate from within the jet core ($\theta \lesssim 20^\circ$).
 
\subsection{Polar outflow}

%Figure \ref{fig-energies} shows the total rotational, internal, and magnetic energy within a box of side length 200\,km. The magnetic energy saturates at $\approx\!10^{51}$\,erg at about 35--40\,ms after merger, when a quasi-stationary jet and accretion disk have been established. The total internal energy remains approximately constant during our simulation run, while the rotational energy slowly decreases, following a linear trend. Indeed, rotational energy is tapped by the magnetic field and the non-axisymmetric density modes to release the magnetized jet and the spiral wave winds.

%\begin{figure}[tb!]
%  \centering
%  \includegraphics[width=\columnwidth]{Emag_vs_t.pdf}
%  \caption{Internal, magnetic, and rotational evolution contained in a box of radius 100 km.}
%  \label{fig-energies}
%\end{figure}

\begin{figure}[tb!]
  \centering
\includegraphics[width=1\columnwidth]{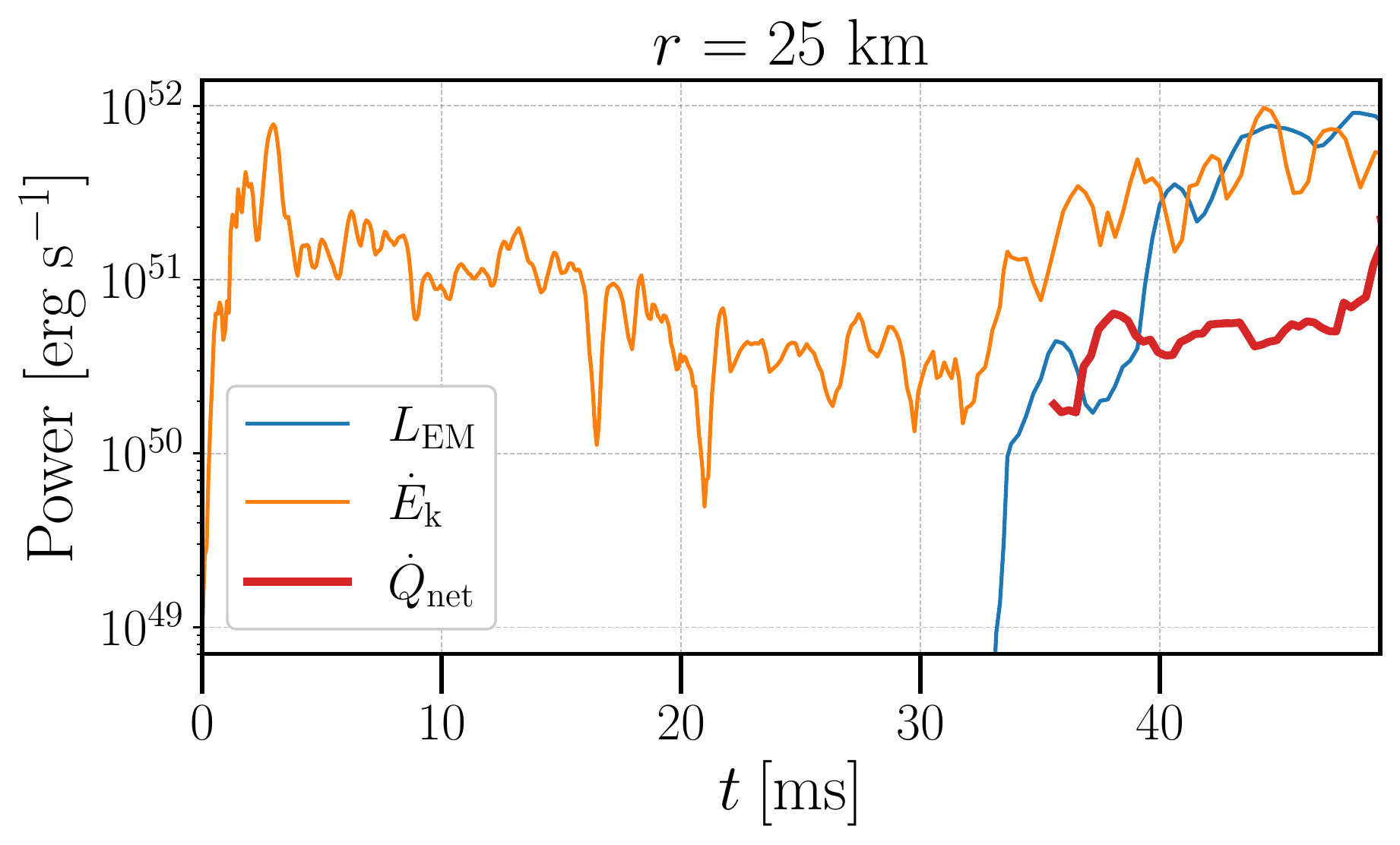}
  \caption{Various luminosities extracted in the polar region $\theta \le 30^\circ$. Shown are the electromagnetic (Poynting) luminosity $L_{\rm EM}$ and the kinetic power $\dot{E}_{\rm k}$ extracted at a radius of 25\,km, as well as the total absorbed neutrino power $\dot{Q}_{\rm net}$ in the corresponding gain layer (a volume between $r=6$\,km and $r=40$\,km). %Dashed lines indicate a linear  extrapolation at early times where we do not have available data for this quantity.
  }
  \label{fig-lum}
\end{figure}

Figure \ref{fig-lum} reports several luminosities in the polar region where outflows dominate over accretion in the immediate vicinity of the stellar surface ($\theta \lesssim 30^{\circ}$). Prior to the jet emergence $t\lesssim 30$\,ms, the kinetic power of the outflow roughly equals the total power absorbed by neutrinos, as expected for a neutrino-driven wind. As the jet emerges the kinetic power rises by more than an order of magnitude and approaches the Poynting luminosity of the emergent jet, which extracts rotational energy from the rotating remnant NS.

\begin{figure}[tb!]
  \centering
   \includegraphics[width=\columnwidth]{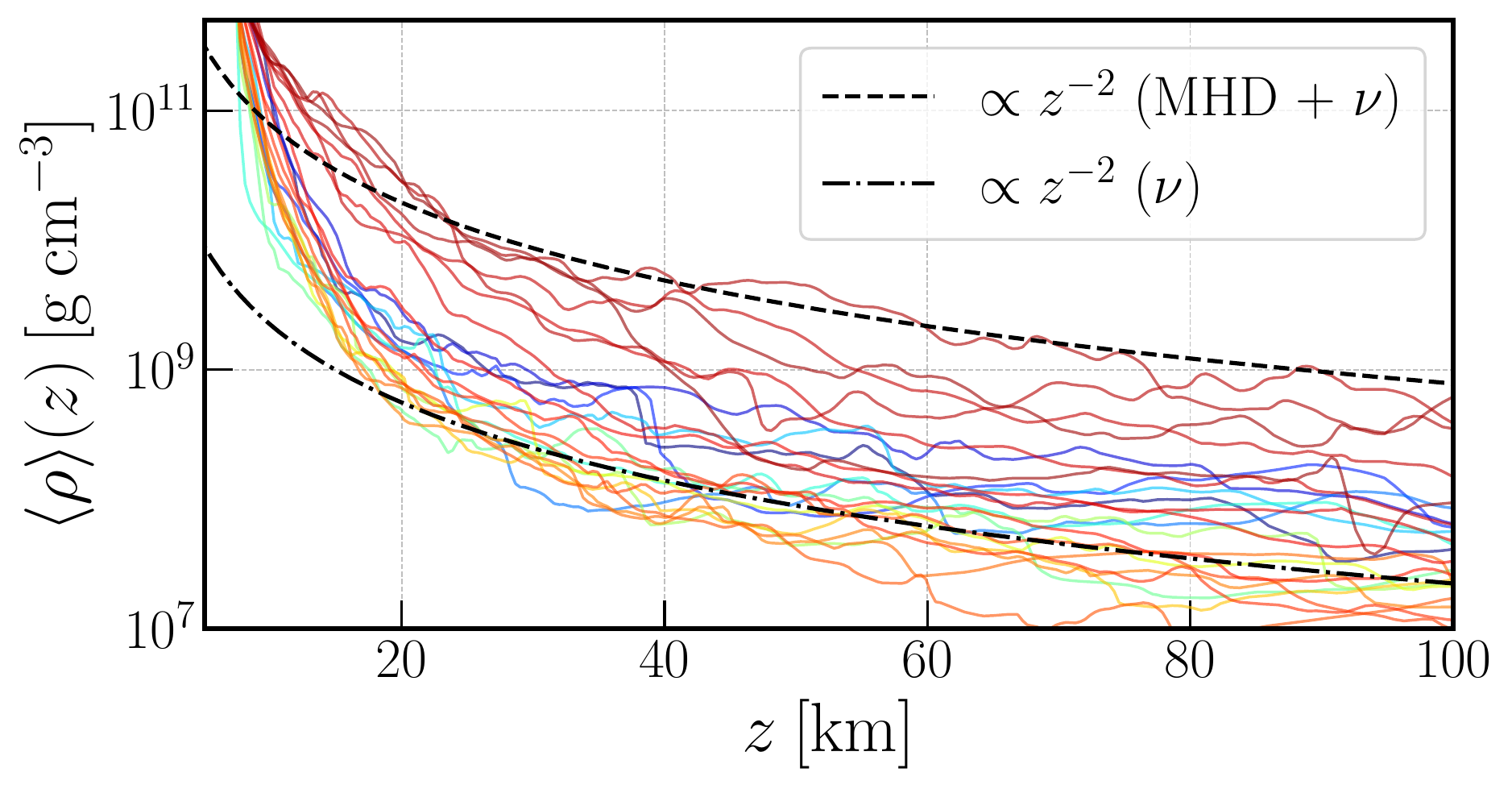}
  \caption{Rest-mass density profiles along the rotational (jet) axis, obtained by averaging over a cylindrical volume $\varpi_{\rm cyl} < 12$\,km, for different times with a frequency of $\approx 1.5$ ms. The blue (early times; neutrino-driven wind) to red (late times; neutrino and magnetically driven outflow) solid lines show the emergence of a stationary ($\propto z^{-2}$) wind profile in both wind regimes. The associated mass-loss rate increases by an order of magnitude as the wind becomes strongly magnetized.}
  \label{fig-magwind-profiles}
\end{figure}

Figure \ref{fig-magwind-profiles} shows rest-mass density profiles along the jet axis at various epochs. Prior to and after the emergence of the jet a $\rho \propto r^{-2}$ wind profile is established as expected from mass conservation for a steady state wind with mass-loss rate and velocity set in the gain region close to the NS surface (see the main text).

The magnetized polar wind with enhanced mass-loss rate of $\dot{M}\approx 1 \times 10^{-2}M_\odot\,{\rm s}^{-1}$ (enhanced by approximately one order of magnitude relative to the prior purely neutrino-driven wind), a poloidal field strength of $\sim\!{\rm few}\times 10^{14}$\,G (Fig.~\ref{fig-B-structure}), a neutrino luminosity of $L_\nu \sim {\rm few} \times 10^{52}$\,erg\,s$^{-1}$, mass-averaged speed $\langle u \rangle \approx c \sigma^{1/3}$ ($\sigma \approx 0.1$; see the main text), and mass-averaged $Y_e\approx 0.3-0.4$ is in broad agreement with the 1D wind solutions of Ref.~\cite{metzger2018Magnetar}.

\subsection{Kilonova light curves from polar and disk winds}

\begin{figure}[tb!]
  \centering
\includegraphics[width=1\columnwidth]{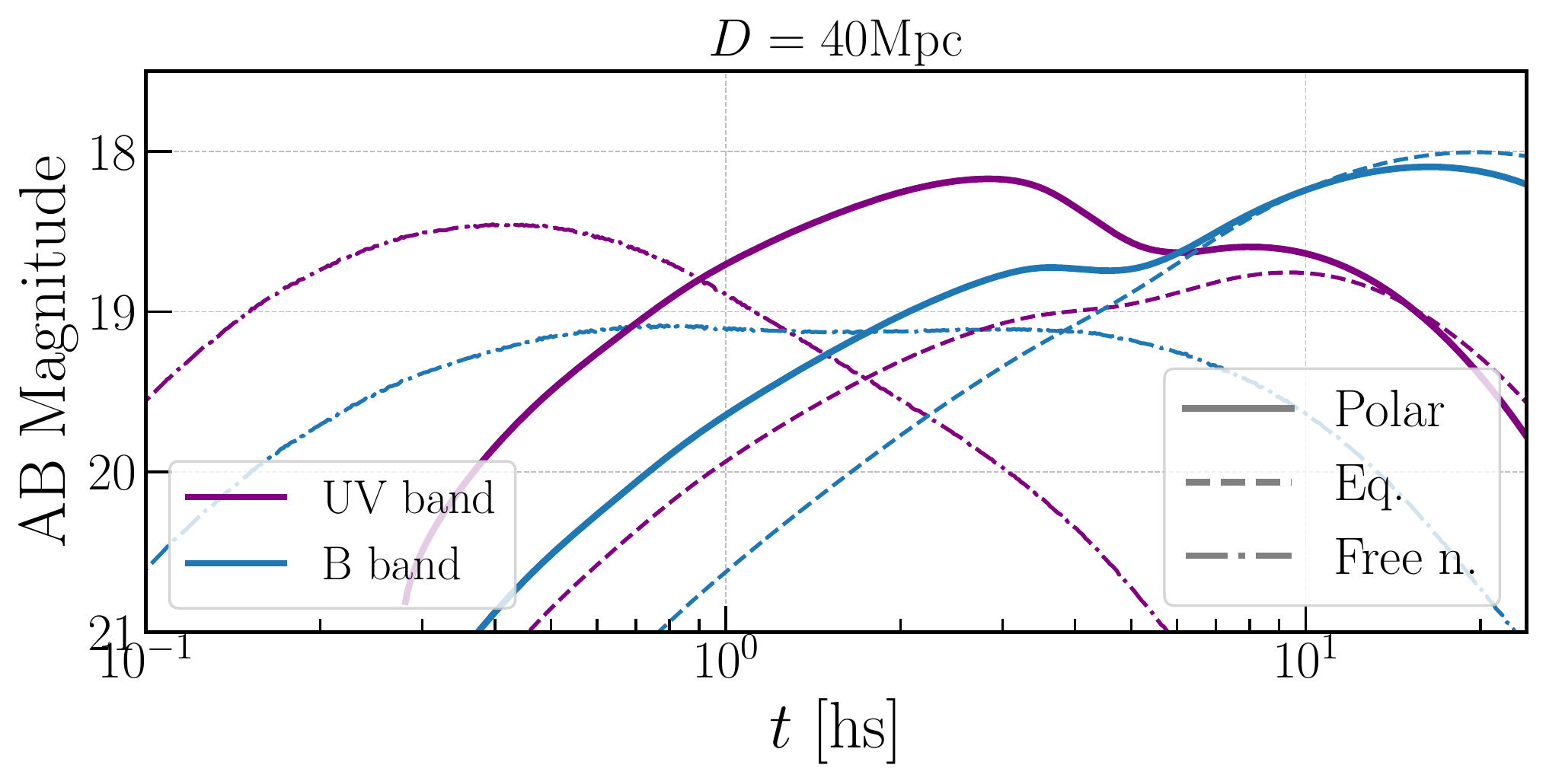}
  \caption{Kilonova light curves in the UV/B bands generated by the post-merger ejecta, viewed along the polar axis (thick lines) and the equatorial plane (dashed lines) at a distance of 40\,Mpc. Also shown for comparison is the kilonova precursor signal due to free neutron decay in fast dynamical ejecta (dot-dashed lines) as computed for this merger setup in CS23.}
  \label{fig-kn-angle}
\end{figure}

Figure~\ref{fig-kn-angle} illustrates the contributions of polar outflows versus disk outflows to the kilonova emission. When observed near the polar axis, the fast polar outflows associated with the jet can greatly boost the emission by an order of magnitude in the UV and blue bands on timescales of a few hours after merger. Dissipation of magnetic energy into heat not considered here may further enhance the emission. Also shown are the kilonova precursor signals due to free neutron decay in fast dynamical ejecta computed in CS23. The polar outflow component peaks somewhat later than the $\lesssim\!1$\,h peak timescale of the neutron precursor, but with similar or higher magnitudes than the neutron precursor. Both emission components strongly overlap in time and create a prolonged precursor signal $\lesssim\!1-\text{few}$\,h.

For calculating our light curves, we use the geometrical multi-angle axisymmetric kilonova approach with the detailed %of Ref. \cite{perego2017AT2017gfo}, combined with the more realistic 
heating rates of Ref. \cite{hotokezaka2020Radioactive} as implemented in CS23. We calculate the $\beta$-decay, $\alpha$-decay, and fission heating rates taking into account elements up to $A=135$ and abundance distributions consistent with our nuclear reaction network calculations. We use opacity values consistent with the ones obtained in Ref.~\cite{banerjee2020simulations} for the early stages ($\approx\!0.1-1$\,day) of a lanthanide-free kilonova. In particular, as a minimal assumption \rvw{and similar to \cite{metzger2015Neutronpowered}}, we use $\kappa = 0.5$\,cm$^2$g$^{-1}$ for $v > 0.2c$, corresponding to the jet component \rvw{with opacity similar to electron-scattering}, and $\kappa = 10$\,cm$^2$g$^{-1}$ for $v < 0.2$, corresponding to the disk wind \rvw{with mostly lanthanide-free, d-block elements}. We have also checked that our results are largely insensitive to using a more detailed, $Y_e$-binned opacity prescription \rvw{(e.g., Ref.~\cite{wu2021Radiation})} as employed by CS23.

\bibliographystyle{apsrev4-2}
\bibliography{ms}% Produces the bibliography via BibTeX.

\end{document}